# Artificial Intelligence in Science: Returns, Reallocation, and Reorganization

Moh Hosseinioun[1,2], Brian Uzzi[1,2], Henrik Barslund Fosse[3]

[1] Kellogg School of Management, Northwestern University
[2] Northwestern Institute on Complex Systems, Northwestern University
[3] Novo Nordisk Foundation

## Abstract

Investment in artificial intelligence (AI) has grown rapidly, yet its returns to scientific research remain poorly understood. We study how AI reshapes the production of science using a comprehensive dataset of research proposals submitted to a large international funding agency, including both funded and unfunded projects. Combining keyword extraction with large language model classification, we identify the presence, type, and functional role of AI within each proposal and link these measures to detailed budget allocations, team structure, and subsequent publication outcomes. We find that, in the short run, AI adoption is associated with modest improvements in scientific outcomes concentrated in the upper tail. Instead, its primary effects arise in the organization of research: AI-enabled projects reallocate resources toward human capital, involve larger teams, and undertake a broader set of tasks. These patterns are consistent with a reorganization of the scientific production process rather than immediate efficiency gains, in line with theories of general-purpose technologies. Task-level analyses further show that activities expanded in AI-enabled projects, particularly ideation and experimentation, are increasingly compatible with large language model capabilities, suggesting potential for future productivity gains as these technologies mature.

## Introduction
Investment in artificial intelligence (AI) research has surged dramatically, with 2024 spending alone far exceeding $79 billion (Nestor Maslej et al., 2025). Yet its returns to scientific research remain poorly understood. AI technologies, widely recognized as instances of general-purpose technologies (GPTs), have the potential to reshape the inputs, processes, and outcomes of research, from how resources are allocated, to how discovery is conducted, to the scientific advances that are ultimately produced (Hao et al., 2023; Duede et al., 2024; Gao & Wang, 2024; Hao et al., 2026; Bresnahan & Trajtenberg, 1995; Brynjolfsson et al., 2021). AlphaFold illustrates this potential transformation by automating key elements of protein structure prediction and reshaping the process of protein discovery. By algorithmically predicting protein structure, it reduces reliance on costly equipment (inputs), restructures discovery workflows toward computation (processes), and has been shown to improve and accelerate specific forms of scientific output (outcomes) (Eisenstein, 2021; D. T. Jones & Thornton, 2022; Kusumegi et al., 2025).

Such transformations across the three major dimensions of scientific production may lower barriers to entry and broaden participation in complex domains, while potentially reducing time and cost constraints. However, such changes may also concentrate advantage. Access to frontier models and the resources required to train them remain highly uneven, potentially reinforcing winner-take-all dynamics. Greater model uncertainty and variability may also require. Nonetheless, AlphaFold, while illustrative, is only one example among many AI methods being developed and adopted across the scientific ecosystem. As scientists consider how to incorporate AI into their research pipelines, and as funding agencies deliberate between AI-enabled and conventional research, the return on investment (ROI) to AI remains uncertain: to what extent do AI models replicate the scale and nature of AlphaFold's effects?

We address this question using a comprehensive dataset of research proposals submitted to one of the largest international health, safety and biomedical funding agencies over the past decade. Leveraging the full set of both funded and unfunded proposals, we apply large language models to systematically identify AI methods mentioned in proposal text and classify their functional roles. This allows us to capture heterogeneity in both the types of AI used and their roles in the research process (Hosseinioun & Tafti, 2023; Lin & Maruping, 2025). Leveraging access to the stage when research ideas are conceived and their execution planned, we examine key project characteristics, including funding likelihood, project duration, and requested budgets, to characterize the inputs and scale of AI-enabled research. Further, we examine outcomes by linking funded proposals to subsequent publications to estimate the scientific returns to AI adoption. Consistent with empirical regularities and GPT theory, we expect a period of latency during which reorganization and learning must take place before tangible outcomes of AI are observed (Brynjolfsson et al., 2021; David, 1990). Linking our algorithmic classifications to detailed budget information on proposals, we provide, to our knowledge, one of the first large-scale examinations of how AI reshapes the allocation of research resources.

We find modest short-run scientific returns despite partially higher investment. AI projects are slightly longer, require similar total budgets, and yield only marginally higher scientific output (Duede et al., 2024). However, AI usage is primarily associated with the reallocation of resources within the research process. Projects that use AI reallocate funds away from equipment and operational expenses toward human capital—particularly salaries—and are undertaken by larger teams. Our findings suggest that, in the short term, AI primarily reorganizes scientific production, consistent with broader patterns observed in the economy, characterized by a latency of tangible outcomes (Brynjolfsson et al., 2021).

A central contribution of this study is its focus on grant proposals, where institutional selection and individual choices jointly shape scientific practice. This approach complements prior work based on publications, which primarily captures realized outcomes rather than ex ante choices. Methodologically, we distinguish between different classes of AI (e.g., statistical learning versus deep learning) and their functional uses (e.g., data collection, analysis, or background reference), allowing us to isolate meaningful adoption from incidental mentions. Together, these distinctions allow us to map AI use onto the tasks that constitute scientific work, clarifying both how AI reshapes research processes and where future productivity gains may arise.

# Analysis

## Measuring AI in research proposals

We develop a multi-stage pipeline to identify and characterize the use of artificial intelligence (AI) in research proposals, allowing us to distinguish both the type of AI adopted and its role in the research process. Candidate AI-related keywords are extracted using a two-pronged approach. In the first step, we apply a combination of curated regular expressions and dictionary-based matching to extract established AI-related keywords (Bianchini et al., 2022; Duede et al., 2024; Dunham et al., 2020; Gao & Wang, 2024; Hao et al., 2026). In the second step, we identify less well-known algorithms, including in many cases novel methods proposed within the project, using contextual information from proposal text interpreted by an LLM. These keyword instances are paired with their surrounding sentences and passed to two sets of large language models (LLMs), which classify each keyword–sentence pair along two dimensions: (i) the architectural class of the method and (ii) its functional role within the proposed research. Fig. 1a depicts this pipeline.

The architectural classification distinguishes between modern AI methods (e.g., neural networks, deep learning, and generative models), statistical machine learning approaches, analytics and data science techniques, and domain-specific algorithmic methods that do not map cleanly to a general architecture. his delineation is important for two reasons. First, different architectural lineages within the broad umbrella of AI may differ in their capacity to reshape the scientific pipeline. Second, the latency of effects predicted by GPT theory may vary across algorithm classes because these technologies are at different stages of their technological life cycles. Our functional classification assigns each instance to one or more use cases within the research process, including ideation, data collection, data processing, analysis, experimentation, inference, validation, automation, application, education, or a benchmark or background reference. This approach allows us to distinguish between substantive use of AI within a project and incidental or referential mentions, ensuring that our measures capture functional adoption rather than simple textual presence. To that end, passing references are excluded from the corpus and our subsequent analyses.

Using these classifications, we construct proposal-level indicators of AI adoption by algorithm class and functional usage. Because proposals may contain multiple methods and serve multiple functions, these measures are non-mutually exclusive. We aggregate these measures over time to characterize patterns of AI adoption and use across the scientific ecosystem (Fig. 1b-c).

Adoption patterns across algorithm classes (Fig. 1b) reveal the rapid diffusion of modern AI, particularly following the release of BERT in 2019 and of AlphaFold and GPT-3 in 2020. A second surge in adoption is seen in 2023, with the introduction of ChatGPT. Breaking adoption down by usage categories further reveals growth in ideation- and experimentation-related uses of modern AI (see Methods for category definitions). SI Figure 4 and SI Figure 5 show that these patterns are robust to alternative aggregation based on keyword counts. Both aggregation methods show that, whereas other algorithm groups are used primarily for data analysis (such as pattern detection, classification, or prediction without scientific interpretation) modern AI is increasingly used for tool and model development (Algorithmic outcomes, automation, model validation), as well as for more complex tasks such as ideation and experimentation. In the remainder of the paper, we focus on modern AI in proposals and discuss other algorithm groups only when their effects differ materially from those of modern AI. Cross-algorithm analyses are presented as supplementary information.

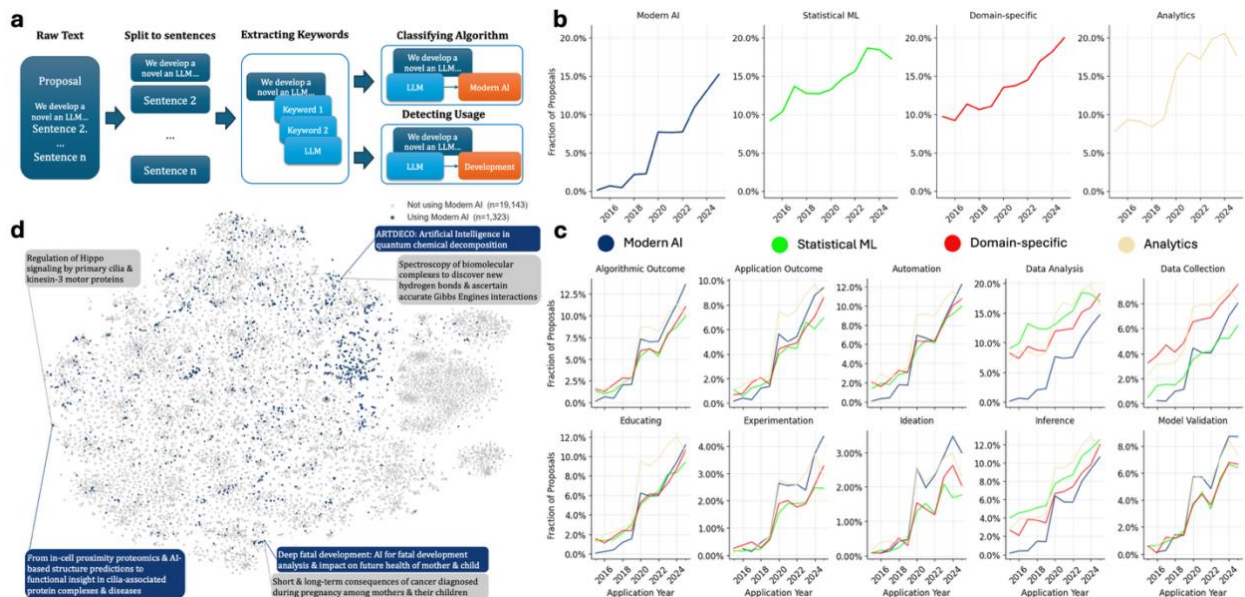

*Figure 1. Identifying AI methods in research proposals and patterns of adoption and use.* (a) Schematic of the multi-stage identification pipeline. Candidate AI-related terms are first extracted using curated regular expressions and dictionary matching, then supplemented with additional algorithmic terms inferred from proposal text using LLMs. Each keyword–sentence pair is subsequently classified by LLMs according to the architectural class of the method and its functional role in the proposed research. (b) Adoption patterns across four algorithmic categories: modern AI (including neural networks, deep learning, and generative models), statistical machine learning, analytics, and domain-specific algorithmic methods. Values show the fraction of proposals containing each category in a given year. (c Adoption patterns across major usage categories, including ideation, data collection, data analysis, model development, and applied systems, excluding incidental references to prior work. Because proposals may contain multiple methods and multiple uses, categories are non-mutually exclusive. (d) Semantic similarity network linking proposals that use modern AI to semantically similar proposals that do not use algorithmic methods, illustrating comparisons between projects addressing closely related scientific problems with different research approaches.

## Descriptive differences in scientific and economic characteristics

We first examine the distribution of key scientific and economic characteristics across AI-enabled proposals, defined as those using methods classified as modern AI, and matched non-AI proposals, defined as those using no detected algorithmic methods. Scientific outcomes are measured using publication-based indicators, including the number of publications, journal impact factors, citation counts, and authorship characteristics. Economic inputs and project structure are captured using proposal-level measures of project duration, team size, total requested budget, and the allocation of expenditures across major budget categories.

These comparisons provide a descriptive overview of how AI-enabled projects differ from conventional research along multiple dimensions of the scientific production function. To further address confounding beyond regression-based controls, we also use proposal text to construct a semantically matched sample of AI-enabled and conventional proposals. This matching procedure allows us to compare projects that address closely related scientific problems while differing in their use of modern AI. We visualize the resulting similarity structure using a two-dimensional -SNE representation in Fig. 1d, highlighting matched pairs of AI-enabled and conventional proposals. This approach provides a qualitative benchmark for comparability and motivates our subsequent analyses.

These analyses provide empirical comparisons of distributions based on matching each AI-enabled proposal to the five most semantically similar proposals in which we detect no algorithm use (Fig. 2a,b). We then estimate regression models in which the outcomes of interest are related to indicators of modern AI use while controlling for relevant proposal-level characteristics, including applicant demographics (age, gender, and prior experience), project length, team size, and textual similarity to contemporaneous proposals. All models also include year and domain fixed effects to account for temporal trends and field-specific differences in research practices. We also implement alternative specifications using continuous and categorical measures of AI intensity (based on the number of AI-related keywords) and regression based on matched samples derived from the semantic pairing procedure. Although unobserved confounding cannot be ruled out, the consistency of results across these specifications supports the robustness of the empirical patterns (see SI Figure 8, SI Table 7, and SI Table 8).

The availability of both funded and unfunded proposals enables counterfactual analyses of both funding decisions and AI adoption. To model funding likelihood as a function of AI use, we estimate logistic regressions, controlling for proposal characteristics, applicant attributes, and similar fixed effects. These models provide estimates of how AI-enabled proposals differ in their funding probability relative to comparable non-AI proposals.

We also incorporate funding status into subsequent regressions to separate the association between AI adoption and project characteristics from selection into funding. This approach enables us to assess whether differences in outcomes reflect the use of AI or the selection of AI-enabled proposals by funding agencies.

## Modern AI use and research inputs

We begin by comparing input factors across AI-enabled and conventional research using both empirical distributions (Fig. 2b) and regression models (Fig. 2d). The outcomes in this section are funding status, project duration, and total requested budget. Although AI-enabled and conventional projects do not differ systematically in total requested budget or funding likelihood, projects that use modern AI are significantly longer. Despite posing a contrasts with the case of AlphaFold, this pattern is consistent with early-stage adoption in a health and biomedicine context, where computational methods may complement rather than replace existing experimental workflows.

## Modern AI use and research processes

To examine how AI adoption is associated with the organization of research processes, we analyze team size and the allocation of proposal budgets across major expenditure categories in Fig. 2b, d. We construct measures of the share of total budget allocated to human capital

(e.g., salaries), equipment and infrastructure, operational costs, and administrative and overhead expenses. The regression models relating these variables to modern AI use follow the same set of controls and fixed effects described above.

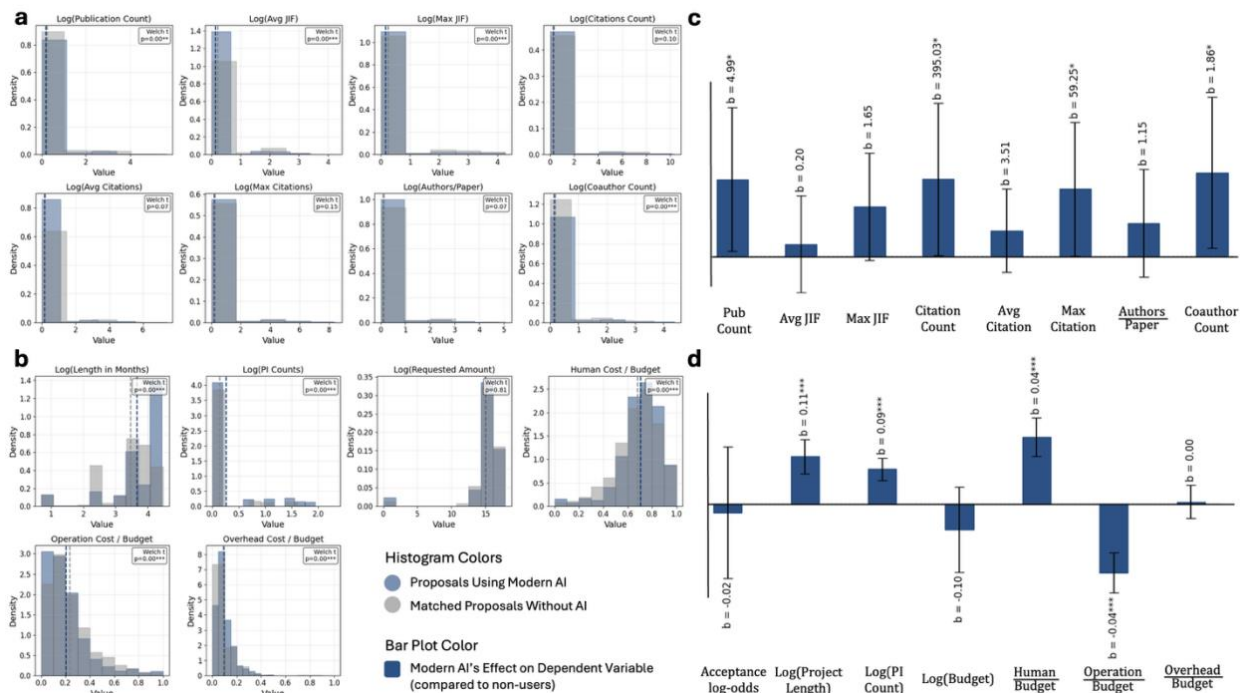

*Figure 2. Scientific and economic characteristics of modern AI-enabled versus conventional research proposals.* Panels (a–b) show empirical distributions for matched samples of proposals using modern AI and semantically similar proposals that do not use algorithmic methods. Panel A reports scientific outcome measures, including publication counts, journal impact factors, citation counts, and authorship characteristics. Panel B reports economic and organizational measures, including project duration, team size, total requested budget, and the shares of budget allocated to human capital, equipment and operations, and administrative or overhead costs. Panels (c–d) depict regression estimates for scientific and economic outcomes associated with modern AI use, conditional on proposal and applicant characteristics, textual similarity to contemporaneous proposals, and year and domain fixed effects. Scientific outcome regressions are estimated for funded proposals linked to subsequent publications; economic and funding regressions are estimated on the full proposal sample. Bars indicate coefficient estimates with 95% confidence intervals. Bar lengths are scaled as a function of coefficients scaled by the variation in the DV.

Both the matched-sample distributions and the regression estimates indicate that modern AI is associated with a reallocation of resources away from equipment and operational costs and toward human capital expenses. This shift coincides with the larger teams involved in AI-enabled proposals.

To further characterize shifts in resource allocation, we decompose aggregate budget categories into more granular components, including personnel, training, and consulting (human capital), equipment, materials, data and computation, experimental activities (operational costs), and administrative and indirect expenditures (overhead). We compare these detailed budget categories across AI-enabled and non-AI proposals, reporting average levels and associated confidence intervals in Fig. 3. The decomposition shows that salaries drive the greater fraction of human capital costs in AI-enabled projects. At the same time, AI-enabled proposals appear to reduce operational costs by lowering general operations, experimental, and material expenses.

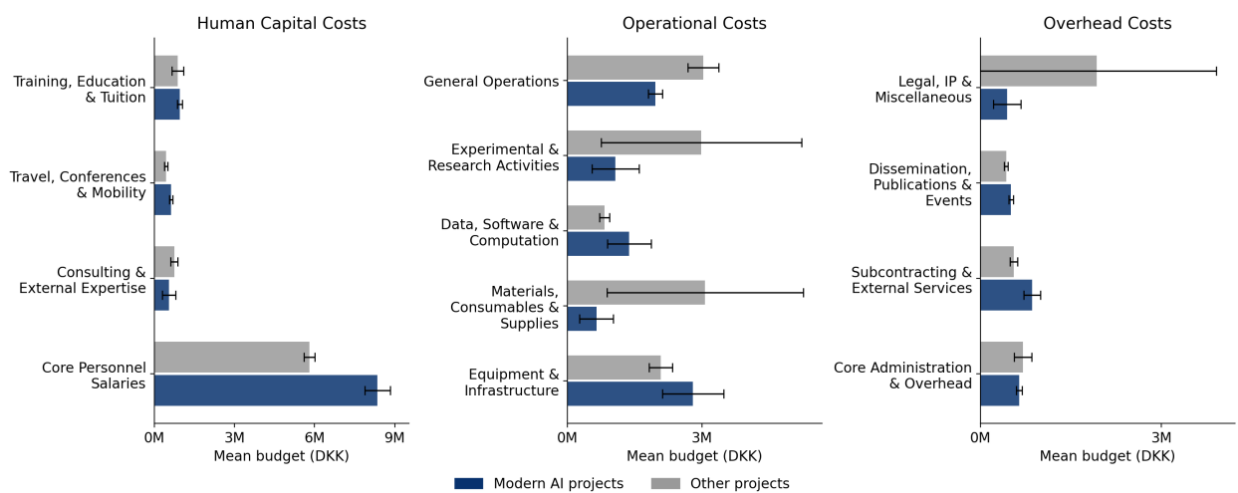

*Figure 3. A Decomposition of the Aggregate Budget Categories into more Granular Components.* These granular components including personnel, training, and consulting (human capital), equipment, materials, data and computation, experimental activities (operational costs), and administrative and indirect expenditures (overhead). Salaries drive the greater fraction of human capital costs in AI-enabled projects. At the same time, AI enabled proposals appear to save operational costs by lowering general operations, experimental, and material costs.

## Modern AI use and scientific outcomes

We next examine the relationship between AI adoption and scientific outcomes in Fig. 2a and 2c. For funded proposals, we link project records to subsequent publications and construct measures of research output and impact, including publication counts, citation-based metrics, and journal-level indicators.

The empirical distributions compare AI-enabled proposals with proposals that do not use any algorithmic methods. In the empirical distributions, unfunded proposals in each group are assigned zeros for these outcome measures. We further estimate regression models that relate these outcomes to AI adoption, conditioning on all previously described project characteristics and explicitly controlling for funding status. These specifications allow us to assess the association between AI use and downstream scientific performance, holding constant observable differences across projects.

Whereas the empirical distributions do not sharply distinguish AI-enabled from other proposals, the regression analyses reveal differences in several upper-tail scientific outcomes, including maximum journal impact factor, maximum citations, and publication counts. A significantly higher number of authors on the publications resulting from funded proposals is consistent with the larger teams involved in AI-enabled projects.

## Task composition and project complexity

An unexpected empirical observation is the greater length of AI-enabled projects, which contrasts with the anecdotal case of AlphaFold. To examine the mechanisms underlying differences in project length, we analyze the task composition of proposals using a taxonomy derived from the Occupational Information Network (O*NET). We map proposal text to a set of task descriptors (Meisenbacher et al., 2025) and construct a measure of the number of tasks associated with each project.

We compare the distribution of tasks across AI-enabled and non-AI proposals and examine how task counts relate to project duration (Fig. 4). AI-enabled proposals contain more tasks on average, and task count is positively associated with project length regardless of AI use.

In Supplementary Information (SI Figs. 9-10), we further analyze differences in the prevalence of specific tasks, ranking them by the difference in frequency between AI-enabled and conventional proposals. This approach reveals that AI adoption is associated with an expansion of activities within projects, rather than with the substitution of computational tasks for activities typically performed in non-AI proposals. This pattern contrasts with the case of AlphaFold, where lengthy exploratory experimentation was partly replaced by computational prediction. In contrast, most AI-enabled proposals appear to undertake the activities of non-AI proposals in addition to their own computationally intensive activities, potentially reflecting the precautionary requirements of research in the health-related context of our data.

The finding that AI-enabled projects are longer raises the question of whether future advances in AI may alter this pattern by improving the productivity of research tasks. The task-level decomposition of proposals allows us to examine where such gains may plausibly arise. In particular, recent evidence from knowledge-intensive production suggests that some components of the research workflow may be more amenable to automation or augmentation by LLMs than others (Dell'Acqua et al., 2026; Hoffmann et al., 2024).

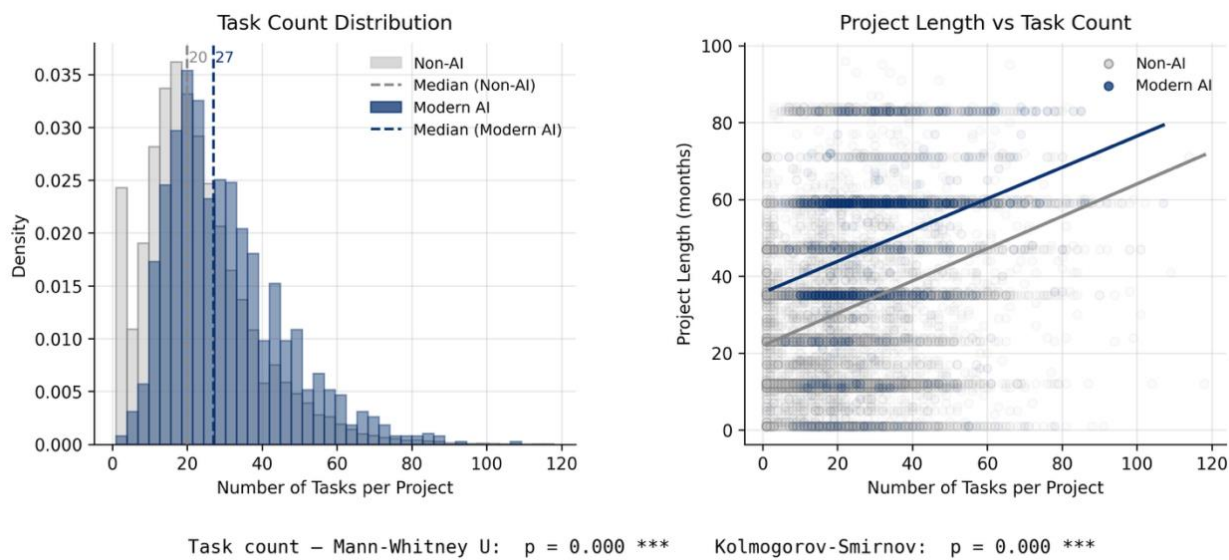

*Figure 4. Task scope and structure of Modern AI versus non-AI research proposals*. The left panel shows the density distributions of the number of distinct tasks per project for proposals classified as Modern AI (blue) and non-AI (gray). Dashed vertical lines indicate group medians. The right panel shows the relationship between task count and project duration (months) for the same two groups, with OLS trend lines overlaid. Each point represents one proposal. Statistical differences in task-count distributions were assessed by two-sided Mann-Whitney U and Kolmogorov-Smirnov tests (statistics reported below panels). Tasks were extracted from proposal texts using a 80% confidence threshold following Meisenbacher et al. (2025).

To assess this possibility, we map extracted proposal tasks to externally validated measures of LLM exposure. Specifically, we use recently released task-level exposure scores developed by Anthropic, which quantify the extent to which a task is compatible with or susceptible to LLM capabilities (Maxim Massenkoff & Peter McCrory, 2026). We aggregate these scores to the proposal level to construct a measure of the overall exposure of project activities to LLMs. We then compare average exposure across AI-enabled and non-AI proposals submitted after 2022, as well as across different usage categories of AI within proposals (Fig. 5).

These exposure patterns suggest that the potential for productivity gains varies systematically across research activities. Tasks associated with ideation and experimentation exhibit the highest exposure to LLM capabilities, indicating that these stages of the research process may be particularly amenable to future efficiency improvements. By contrast, tasks that are less compatible with current LLM capabilities may continue to drive the cost and length of AI-enabled projects. Although we do not directly observe the integration of LLMs into project execution, the alignment between emerging modern AI use cases and high-exposure tasks suggests one pathway through which advances in AI may translate into future improvements in the productivity and organization of scientific work (Brynjolfsson et al., 2025).

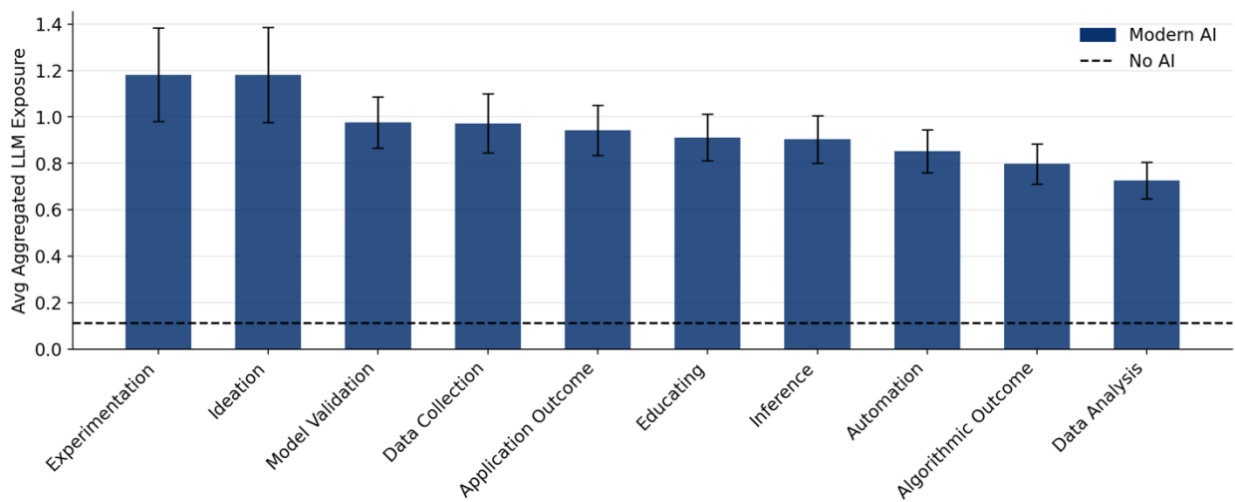

*Figure 5. Aggregated LLM Exposure across Usage Categories of AI-enabled Proposals.* To estimate the LLM exposure for each proposal, we first match extracted tasks to the newly released Anthropic LLM exposure scores. These exposure scores are then summed at the level of a proposal, aggregated LLM exposure, to measure the amenability of the proposal activities to be performed by LLMs. SI Figure 12 expands this comparison to all algorithm categories. All exposure estimates limit the sample to proposals after 2022, when ChatGPT was released.

# Conclusion

A growing empirical literature has examined the role of artificial intelligence (AI) in science, with three complementary research streams emerging. First, a set of studies analyzes the incorporation of AI methods into scientific research itself, typically using large corpora of publications to track adoption patterns and their relationship to novelty, productivity, and the direction of inquiry (Bianchini et al., 2022; Duede et al., 2024; Hao et al., 2023, 2026). These studies document the diffusion of AI across fields and highlight heterogeneity in both adoption and its relationship to scientific productivity. Second, a growing literature focuses on the use of LLMs in the writing and reporting of science, examining how these tools affect the novelty, framing, and success of scientific dissemination and funding (Liang et al., 2025; Qian et al., 2026). Third, experimental studies evaluate the impact of AI on scientific and innovative tasks more directly, offering counterfactual but restricted evidence that generative AI can improve average performance while altering the distribution of outcomes and the nature of creative problem-solving (Boussioux et al., 2024; Dell'Acqua et al., 2026).

Our analysis complements all three streams. We expand the unit of observation beyond published research to the entire corpus of research proposals received by one of the largest international biomedical and health funding agencies. By examining both funded and unfunded proposals, we provide a large-scale counterfactual view of AI adoption that captures both successful and unsuccessful scientific efforts. This broader lens allows us to identify patterns of adoption prior to selection and to more fully characterize the relationship between AI use, resource allocation, and downstream outcomes, including the margin of unsuccessful scientific efforts typically unobserved in publication-based data. Our work aligns with the emerging perspective that characterizes AI as a general-purpose tool, operating across multiple stages of research, from hypothesis generation and experimental design, to analysis, and dissemination(Boyko et al., 2023; Chen et al., 2025; Zhang et al., 2025). This view suggests that the impact of AI on science may be best understood not only through its effects on outputs, but through its influence on how scientific work is organized and executed.

Our results point to this very pattern. We find that, in the short run, the adoption of modern AI is associated with first-order gains concentrated in the upper tail of outcomes, and in some cases, increased input requirements, such as longer project durations. Instead, the primary effects of AI appear in the process of scientific production. AI-enabled projects reallocate inputs away from equipment and operational expenditures toward human capital, expand the scope of tasks undertaken, and rely on larger and more diverse teams. In this respect, the effects we document echo the transformation observed in the case of AlphaFold, while also revealing that such changes are not uniformly accompanied by reductions in cost or time (Gil & Moler, 2025).

These findings offer evidence consistent with a broader view of AI as a general-purpose technology that reorganizes, rather than immediately enhancing, the knowledge production process (Agrawal et al., 2026). Paired with the documented growth in the complexity and collaborative nature of science (Jones, 2025), our results provide micro-level evidence of increased coordination through larger teams, an expansion of task scope, and a reallocation of resources toward human inputs. In this sense, we document concentrated and restricted short-run returns that resemble patterns observed in the broader economy, where general-purpose technologies often induce a period of reorganization before productivity gains materialize (Baily et al., 2025; Brynjolfsson et al., 2021; Hosseinioun & Tafti, 2024). The limited first-order gains we observe may therefore reflect a transitional phase characterized by reallocation and reorganization rather than immediate increases in output.

What this transition implies for the future role of scientists, however, remains less clear. Advances in LLMs have expanded the range of tasks that can be partially automated, raising questions about the relevance of human expertise that parallel long-standing discussions in the labor market(Autor & Thompson, 2025; Brynjolfsson et al., 2025). Our finding that resources shift from operational and equipment costs toward human capital expenditures suggest that, at least in the current period, AI complements rather than substitutes for scientific labor. In our setting, this shift reflects both a decline in experimental and operational costs and an increase in salary expenditures, which points to a growing demand for human inputs even as computational tools become more capable. How these changes affect the expertise required of scientists and the long-run evolution of skill acquisition remains an open question (Acemoglu et al., 2026; Garicano & Rayo, 2025).

While these patterns characterize the current phase of AI adoption, our task-level analysis provides insight into where future productivity gains are most likely to emerge. Many of the activities undertaken in AI-enabled health and biomedical research projects, particularly ideation and experimentation, are increasingly compatible with LLM capabilities. As these technologies continue to improve, they may enable efficiency gains in stages of the research process that are currently resource-intensive. In the absence of clear short-term effect on scientific output, adoption decision by scientists may rely closely to risks of shifting to new methods, which likely vary by domain (Hill et al., 2025; Jin et al., 2019).

From an institutional perspective, the absence of large first-order gains in scientific output should be interpreted in the context of these ongoing organizational changes. The reallocation of resources, expansion of task scope, and growth in team size may have second-order implications for the structure of scientific collaboration and knowledge production. Prior work has documented the increasing importance of teams in science and the distinct roles that team size plays in shaping the direction of innovation (B. Jones, 2009, 2025; Wu et al., 2019; Wuchty et al., 2007). Our findings suggest that AI may reinforce these trends through increased reliance on diverse expertise and larger team sizes. The long-run impact of AI on science will, however, depend less on immediate gains in output and more on how it reshapes the organization, coordination, and incentives that govern knowledge production. As both technological capabilities and adoption patterns remain in an early stage, institutional decisions around funding, evaluation, and workforce development will be central in determining how these changes affect the organization of science and the role of scientific labor.

# Method

## Identification and Classification of AI-Related Keywords

We identify AI- and ML-related content in research proposals using a three-stage pipeline designed to capture both established methods and context-specific references. In Stage 1, mentions were detected through two passes: a deterministic scan against a seed lexicon of roughly 200 canonical AI, ML, and analytics terms drawn from four prior studies (Hao et al., 2023; Duede et al., 2024; Gao & Wang, 2024; Hao et al., 2026) (See SI Table 1 for the list of terms), followed by a large language model (Qwen2.5-32B-Instruct(Yang et al., 2024)) applied only to sentences where the initial scan detected no keyword, to capture domain-specific tools and paraphrased references outside the fixed lexicon (See SI Table 2 for the prompt). In our search, we excluded proposals written in a non-English language. Extracted keywords were retained at a confidence threshold of 0.70, chosen because stricter cutoffs produced negligible reductions in false-positive rates while substantially shrinking coverage. A manual review of high-frequency terms removed false-positive patterns, generic acronyms, experimental design terms, and domain abbreviations misclassified as AI tools, yielding 8,762 unique canonical keywords across 71,479 keyword–sentence pairs drawn from 18,896 proposals.

In Stage 2, each keyword–sentence pair was assigned to one or more of eleven mutually non-exclusive research-workflow roles using Meta-Llama-3.1-70B-Instruct(Meta, 2024), with definitions written to reference each specific keyword to reduce ambiguity (See SI Table 3 for the definitions and SI Table 4 for the prompt). These roles distinguish how a tool is used during the research process, including ideation (generating hypotheses or research designs), data collection (acquiring or synthesizing data), data analysis (pattern detection, classification, or prediction without scientific conclusion), experimentation (running trials or simulations), inference (drawing causal or diagnostic conclusions), model validation (assessing performance or generalizability), automation (substituting human effort in a workflow), and benchmarking (contextualizing the tool relative to prior work), from outcome-oriented roles, where the AI is itself what is produced (algorithmic outcome: a trained model or method; application outcome: a deployed tool or platform) or transmitted (educating: taught within a training program or curriculum).

To assess classification reliability, we compared outputs from Llama-3.1-70B with an independent run using Qwen2.5-32B-Instruct. Agreement was moderate overall (Jaccard similarity = 0.41; average per-stage Cohen's κ = 0.31), with higher reliability for well-defined stages such as educating and benchmarking and lower agreement for more interpretive categories such as ideation and experimentation. To mitigate these cases of low agreement, all instances labeled as ideation or experimentation were manually reviewed, relabeled when possible, and removed otherwise.

In Stage 3, each keyword was assigned to one of ten algorithm categories using Qwen2.5-32B-Instruct: Deep Learning and Neural Networks (architectures whose primary structure is a neural network, including convolutional, recurrent, and transformer models); Generative Models (models primarily designed to generate data or predict structural configurations, e.g., GANs, diffusion models, AlphaFold, GPT); Statistical and Classical ML (non-neural supervised and unsupervised methods, e.g., random forests, SVMs, PCA); Reinforcement Learning (algorithms optimizing policies through interaction and reward); Domain-Specific Pipelines (image, text, or biological signal processing pipelines not tied to a specific ML paradigm); Data Science and Analytics (broad, paradigm-agnostic analytic or data-mining workflows); Learning Paradigms (terms denoting training strategies rather than specific models, e.g., supervised learning, transfer learning); Broad ML and Broad AI (non-specific mentions of machine learning or artificial intelligence without further qualification); and Unknown/Domain-Specific (proprietary, opaque, or unidentifiable algorithms). See SI Table 5 for definitions, illustrative examples, and SI Table 6 for the prompt. Assignments with confidence below 0.50 were discarded, and generic terms (e.g., "algorithm") were reassigned to the relevant category through extensive manual review. For less well-known algorithms classified based on contextual information, we assign the majority class and manually review classifications for accuracy. See SI Figure 3 for the most frequently observed keywords in each algorithm categories. For analysis, the ten categories were collapsed into four groups to capture meaningful differences in technological capability and stage of development: Modern AI (deep learning and generative models), Statistical ML, Domain-specific methods, and Analytics.

## Classifying Budget Items

Budget records from the funding institute contain, for each line item, a standardized budget post label, a free-text description, and a year. Budget post labels vary in specificity: some are already precise and directly usable (e.g., *Salary*, *Salary – PhD student*, *Salary – postdoc*), while others are coarse or ambiguous (e.g., *Other*, *Operating expenses*). For the latter, the free-text budget description was used to assign a more informative category. Each unique description was encoded using a multilingual sentence-embedding model (paraphrase-multilingual-mpnet-base-v2 (Reimers & Gurevych, 2019)) and matched by cosine similarity to a set of predefined category labels, designed to resemble the original budget post taxonomy but refined to improve semantic matching, spanning personnel, travel, training, equipment, software, computation, administration, and dissemination. Throughout, the classification scheme and aggregation decisions were validated in consultation with financial and grant relations officers at the School of Management to ensure conceptual coherence with funder accounting conventions.

This classification enables us to measure how AI adoption is associated with the reallocation of resources across key inputs to scientific production. The resulting line items were aggregated to the proposal level across all grant years and organized into a two-level hierarchy. At the intermediate level, items were assigned to thirteen subcategories grouped into three buckets: *Human Capital Costs* (*Core Personnel Salaries*, *Consulting and External Expertise*, *Travel, Conferences and Mobility*, *Training, Education and Tuition*); *Operational Costs* (*Equipment and Infrastructure*, *Materials, Consumables and Supplies*, *Data, Software and Computation*, *Experimental and Research Activities*, *General Operations*); and *Overhead Costs* (*Core Administration and Overhead*, *Subcontracting and External Services*, *Dissemination, Publications and Events*, *Legal, IP and Miscellaneous*). These thirteen subcategories were then collapsed to the three coarse components for the primary analysis. Each component is expressed both as an absolute amount and as a share of total approved budget. Two additional personnel variables were derived: *budgeted team size*, the average number of distinct salary line items per budget year across the grant period, and *principal investigator salary*, the total amount allocated to the main applicant.

## Collecting and Producing Scientific Outcome Variables

Publication records were provided directly by the funding institute for the subset of funded proposals. Bibliometric metadata, including citation counts, author lists, and journal classifications, were retrieved for each publication from OpenAlex. Where OpenAlex returned no match, the journal name recorded in the institute's publication records was used as a fallback. Journal Impact Factors (JIF) were obtained from the Web of Science Journal

Citation Reports and matched to publication records by journal name. If a proposal has no associated publications, all outcome measures are set to zero. This coding ensures that outcome measures reflect both selection into publication and realized research output.

All measures were aggregated to the proposal level. Publication count is the number of publications recorded for a given proposal. Citation count is measured in three ways based on the cumulative cited-by counts from OpenAlex at the time of data collection: total citations summed across all publications, mean citations per publication, and the maximum citations of any single publication. Journal quality is captured by the mean and maximum JIF across all publications associated with a proposal. Count of authors per paper is the mean number of unique authors per publication, where authorship lists were deduplicated within each paper using OpenAlex author identifiers before averaging across publications in a proposal.

### Extracting Tasks from Proposal Text

The activity content of each proposal was characterized to capture the composition and scope of research tasks using the Job Advertisement Analysis Toolkit (Meisenbacher et al., 2025) (JAAT). JAAT is a transformer-based multi-label classifier trained on over 155 million job postings from the National Labor Exchange (NLx) to map unstructured text onto task statements from the O*NET occupational taxonomy. For each proposal, JAAT identified which O*NET task statements (standardized descriptions of discrete work activities) were present in the proposal text. After breaking down each proposal into blocks of 500 words, we applied JAAT at confidence threshold of 80 to extract tasks. The task count variable records the number of distinct O*NET task statements extracted per proposal. This measure allows us to examine whether AI adoption is associated with substitution across tasks or an expansion of task scope within projects.

Raising the threshold from 80% to 90% reduces coverage but leaves the rank ordering of proposals and statistical contrasts between AI and non-AI projects unchanged. We therefore adopt the 80% threshold to maximize coverage.

Two caveats apply. First, JAAT was developed and validated on job advertisements, not research proposals; to the extent that applicants describe their planned activities in task-oriented language, which grant proposals are expected to do, the classifier's representations transfer reasonably, but detection rates may be lower for activities described in highly abstract or theoretical terms. Second, O*NET's task taxonomy was designed to characterize occupational work broadly, not to demarcate research activities specifically. Nonetheless, because O*NET covers a wide range of research occupations (biological scientists, social scientists, physicists, and related roles) at considerable granularity, its task statements offer sufficient resolution to describe and distinguish the activity profiles of research proposals.

# Supplementary Information
## Detecting and Classifying Keywords

The seed lexicon was assembled from four prior bibliometric studies. Terms are listed in SI Table 1 by source after deduplication and punctuation normalization (lowercase, punctuation removed).

*SI Table 1. Seed Lexicon Used in Detecting Algorithmic Keywords*

| Source | Terms |
|---|---|
| General / Core | machine learning, deep learning, neural network, artificial intelligence, reinforcement learning, support vector machine, random forest, GPT-4, data science, computer vision, analytics, big data, data mining, data analysis, data analytics, AI algorithm, AI model, AI system, AI tool |
| Hao, et al. (2026) | gradient boosting, linear discriminant analysis, k-nearest neighbour, k nearest neighbor, k-nearest neighbors, principal component analysis, long short-term memory |
| Jao & Wang (2024) | machine learning, recurrent neural network, convolutional neural network, decision tree, deep learning, reinforcement learning, support vector machine, supervised learning, deep neural networks, pattern recognition, artificial intelligence, natural language processing, computer vision, genetic algorithm, random forest, machine translation, artificial neural network, word embedding, generative adversarial network, extreme learning machine |
| Dunham et al. (2020) | active learning, incremental clustering, adaptive learning, information extraction, anomaly detection, information fusion, information retrieval, associative learning, k nearest neighbor, autonomous navigation, knowledge based system, autonomous system, knowledge discovery, autonomous vehicle, knowledge representation, average link clustering, language identification, back propagation, backpropagation, machine perception, binary classification, bioNLP, multi class classification, boltzmann machine, multi label classification, character recognition, multi task learning, classification algorithm, natural language generation, classification label, natural language processing, clustering method, natural language understanding, complete link clustering, neural network, computer aided diagnosis, object recognition, computer vision, one shot learning, deep learning, pattern matching, ensemble learning, pattern recognition, evolutionary algorithm, random forest, face expression recognition, recommend system, face identification, recurrent network, face recognition, reinforcement learning, feature extraction, scene classification, feature learning, scene understanding, feature matching, self driving car, feature selection, semi supervised learning, feature vector, sentiment classification, feedforward network, single link clustering, feedforward neural network, spatial learning, fuzzy clustering, speech processing, generative adversarial network, speech recognition, gradient algorithm, speech synthesis, graph matching, statistical learning, graphical model strong, handwriting recognition, supervised learning, hierarchical clustering, support vector machine, hierarchical model, text mining, human robot, text processing, image annotation, transfer learning, image classification, translation system, image matching, unsupervised learning, image processing, video classification, image registration, video processing, image representation, weak artificial intelligence, image retrieval, zero shot learning |
| Bianchini et al. (2022) | neural network, neural networks, artificial neural, artificial neural network, deep learning, convolutional neural, convolutional neural network, recurrent neural, recurrent neural network, deep neural, multilayer perceptron, deep neural network, hidden layer, deep convolutional, deep convolutional neural network, long short term memory, hidden layers, restricted boltzmann, auto encoder, generative adversarial, encoder decoder, adversarial network, generative adversarial network, fully convolutional network, convolutional layers, variational autoencoder, adversarial attacks, adversarial examples, variational autoencoders, adversarial perturbations |

The prompt in SI Table 2 was applied to sentences where the regex scan found no match. {*sentence_text*} is replaced with the focal sentence. SI Fig. 1 shows the top 20 keywords observed in the data.

*SI Table 2. LLM Prompt — Stage 1: Keyword Extraction*

> You are an experienced scientist with extensive computational expertise, analyzing scientific research proposals.
>
> The sentence is extracted from a research proposal written by scientists describing their methods and research design.
>
> Task:
> Extract all mentions, as well as explicit, literal mentions of Artificial Intelligence, Machine Learning, or Analytics tools and algorithms.
> Examples include exactly appearing phrases such as 'deep learning', 'machine learning', 'AI algorithm', 'neural network', 'data science', 'GPT-4', 'AlphaFold', 'RosettaFold', 'VeriPB', etc.
> Infer whether a term different from the seed is an algorithm, but include only terms that literally appear in the text.
>
> For each detected mention, provide:
>   - 'raw_span': the exact surface form as it appears in the text
>   - 'canonical_name': standardized version (e.g., 'Deep Learning', 'Random Forest')
>   - 'category': one of {'AI', 'ML', 'Analytics'}
>   - 'rationale': short justification why this is AI, ML, or Analytics
>   - 'confidence': number between 0.0 and 1.0
>
> Return STRICT JSON only, in this exact format:
> {'mentions': [{'raw_span': '...', 'canonical_name': '...',
> 'category': '...', 'rationale': '...', 'confidence': 0.0}]}
>
> Sentence: {sentence_text}

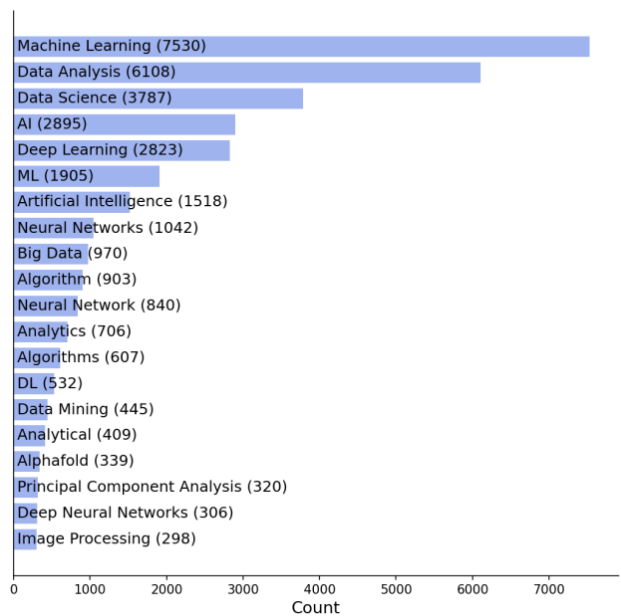

*SI Figure 1. Top 20 Keywords Observed in Proposal Texts*

The usage categories and definitions applied to classify algorithms are shown in SI Table 3.

*SI Table 3. Usage Classes and their Definitions Applied to Classify Algorithms*

| Usage Categories | Definition |
|---|---|
| Ideation | The tool directly generates research ideas, formulates hypotheses, conceptualizes approaches, narrows the scope of a problem, or designs solutions, protocols, or research steps. Excludes general statements about future research or exploration. |
| Benchmarking | The text describes what the tool can or cannot do, or was or was not used for in prior literature, or compares past work against what the proposal does. Excludes statements about how or for what the tool will be used in the focal project. |
| Data Collection | The tool is used to acquire, collect, extract, gather, store, measure, generate, or create data or samples. |
| Data Analysis | The tool is used to analyze data, detect patterns, classify, regress, cluster, predict, estimate, or create descriptive summaries, without drawing scientific conclusions. |
| Experimentation | The tool is used to run experiments, conduct trials, or model simulations. Excludes "experiment" in the sense of exploring or examining. |
| Inference | The tool is used to interpret results, test hypotheses, make diagnoses, draw conclusions, reveal causal relationships or mechanisms, reason about findings, or recommend action or care. |
| Model Validation | The tool undergoes validation, evaluation, or testing to assess its performance, accuracy, reliability, or generalizability. |
| Automation | The tool is used to automate tasks, processes, or workflows, or to substitute humans in certain tasks. |
| Algorithmic Outcome | The tool is the model, system, or algorithm being created, developed, trained, refined, or advanced. Used when the sentence describes why and how a model or algorithm was developed. |
| Application Outcome | The tool is a real-world applied software tool, web platform, application, dashboard, clinical decision support tool, or operational system. |
| Educating | The tool is being taught or disseminated to individuals (e.g., students, researchers), used for training or education, or incorporated into a curriculum. |

The prompt in SI Table 4 was applied per keyword–sentence pair. {keyword} and {sentence} are replaced with the focal term and sentence; {stage_defs} is replaced with the full keyword-specific stage definitions from Table S1. On parse failure, the prompt is retried with the surrounding ±1-sentence context substituted for {sentence}.

*SI Table 4. LLM Prompt — Stage 2: Research Stage Classification*

> Overview: You are an expert annotator, classifying the role played by a tool in a specific part of a research proposal as described by a sentence.
>
> tool: {keyword}
> sentence: {sentence}
>
> Task: Determine which research usage(s) the tool is used for in the SENTENCE above.
> Research usages: {stage_defs}
>
> Rules (follow strictly):
> - Consider ALL usages jointly.
> - Assign ONLY usages that are explicitly mentioned in the sentence.
> - Prefer the most specific usage(s).
> - Do NOT infer unstated or downstream uses.
> - If no usage clearly applies, assign all scores = 0.

> - If the score for a usage is 0, leave its rationale blank.
>
> Scoring guidance:
> - 1.0  = the usage is EXPLICITLY AND DIRECTLY mentioned.
> - 0.75 = the usage is EXPLICITLY AND DIRECTLY stated, but only partially aligned with the definition.
> - 0.5  = the usage is EXPLICITLY OR DIRECTLY stated and only partially aligned with the definition.
> - 0.25 = the usage is IMPLIED or WEAKLY ALIGNED with the definition.
> - 0.0  = the usage is not mentioned, or is misaligned.
>
> Use the FULL score range when appropriate. Reserve 1.0 ONLY when {keyword} is EXPLICITLY AND DIRECTLY mentioned in the defined usage.
> If you are IMPLYING the usage, give 0.25.
>
> Return STRICT JSON only:
> {"scores": {"stage_name": 0.0–1.0},
>  "rationales": {"stage_name": "Brief explanation"}}
> Return ONLY valid JSON. Do not include explanations.

Because no large-scale human annotation of research workflow stages exists for this corpus, we assessed classification reliability by comparing outputs from two independently developed LLMs, Meta-Llama-3.1-70B-Instruct and Qwen2.5-32B-Instruct, applied to the same keyword–sentence pairs under identical prompting conditions. Convergence between two models differing in architecture, scale, and training data provides evidence that classifications reflect genuine, recoverable signal in the proposal text; systematic divergence, by contrast, identifies stages whose linguistic expression in grant proposals is inherently ambiguous and therefore less suitable for downstream analysis.

The two models agreed on a multi-label exact match in 36% of records (Jaccard similarity = 0.41). Per-stage Cohen's κ averaged 0.31, masking substantial heterogeneity: educating was reliably classified by both models (κ = 0.74), as were benchmarking, data collection, and data analysis (κ = 0.40–0.45), while experimentation and ideation showed near-chance agreement (κ = 0.06 and 0.12). Per-stage confusion matrices (see SI Figure 2) corroborated this picture: for nine of the eleven stages, disagreement rates fell below 10%. The two stages with highest disagreement, data analysis (32%) and algorithmic outcome (15%), were driven by false negatives in which Qwen assigned the stage but Llama did not, consistent with Llama's more conservative posture (average precision 0.70, recall 0.29). The two models also differed markedly in coverage: Qwen assigned at least one stage to 94% of records versus 48% for Llama, with the largest gaps in experimentation (Qwen: 6,693 records; Llama: 231) and ideation (Qwen: 5,003; Llama: 462). To mitigate this variation, all instances labeled as ideation and experimentation were manually reviewed, relabeled when possible, and removed otherwise. Off-diagonal counts were small relative to the dominant true-negative quadrant, indicating that both models largely concur on the absence of a stage when it is not expressed.

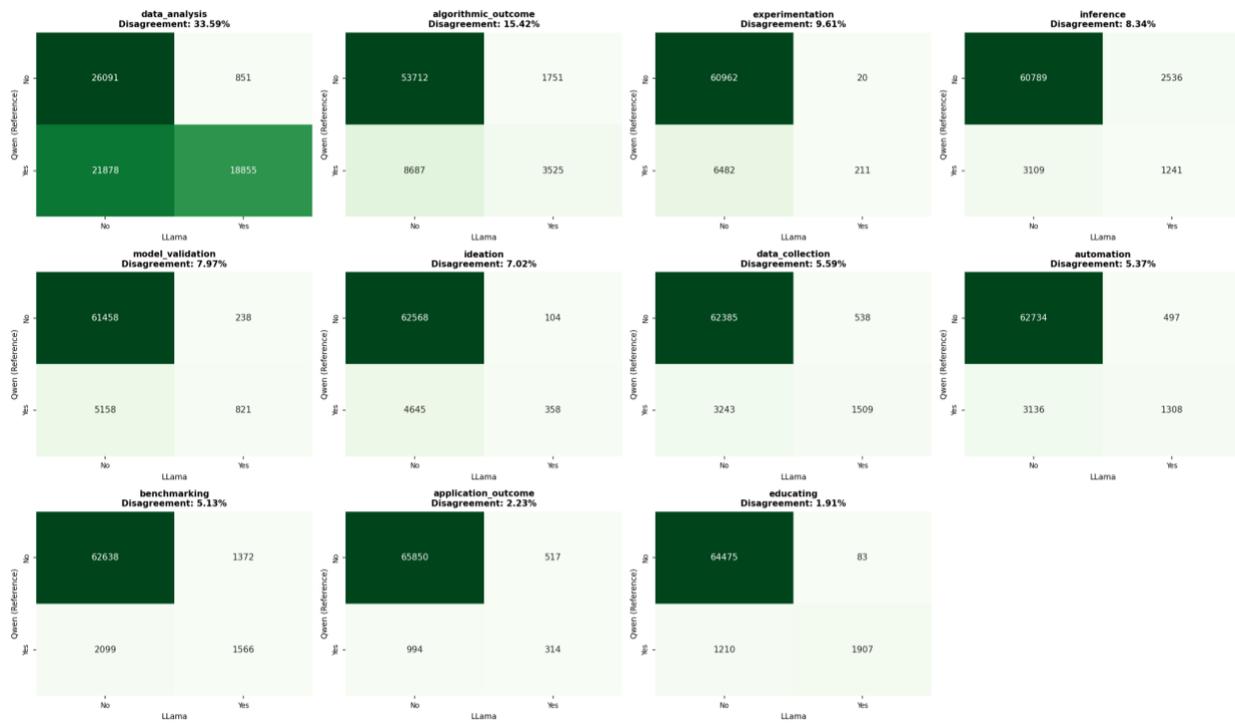

*SI Figure 2. Confusion Matrices Describing Classification of Records for which at Least one Model Assigned the Keyword to a Class.*

SI Table 5 captures the algorithm categories, their definitions and illustrative examples.

*SI Table 5. Algorithm Categories and Illustrative Examples Applied to Categorize Algorithms*

| Category | Definition | Illustrative Examples |
|---|---|---|
| Deep Learning and Neural Networks | Algorithms whose primary structure is a neural network, including shallow, deep, convolutional, recurrent, or transformer architectures. | CNN, RNN, LSTM, BERT, Graph Neural Network |
| Generative Models | Models whose primary purpose is generating new data, learning latent generative structure, or predicting structural configurations. | GAN, Diffusion Models, Variational Autoencoder, AlphaFold, GPT, ChatGPT |
| Statistical and Classical ML | Traditional statistical models and classical machine learning algorithms that do not rely on deep neural networks. | Logistic Regression, PCA, Random Forest, SVM, K-means, Bayesian Optimization |
| Reinforcement Learning and Decision Making | Algorithms optimizing actions and policies through interaction, rewards, or search. | Q-learning, Actor-Critic, Policy Gradient, Monte Carlo Tree Search |
| Domain-Specific Signal, Image, and Text Pipelines | Data processing or analysis pipelines for images, biological sequences, text, or signals that do not rely on a specific ML paradigm. | Image Processing, Image Analysis, Text Mining, Deep Sequencing, Object Recognition |
| Data Science, Data Mining, and Analytics | Broad, non-specific analytic, exploratory, or computational workflows without a single underlying ML paradigm. | Data Science, Data Mining, Analytics, Big Data, NVivo |
| Learning Paradigms and Training Strategies | Terms describing modes of learning or training strategies rather than specific models. | Supervised Learning, Unsupervised Learning, Transfer Learning, Active Learning |

| Broad ML | Broad, non-specific mentions of machine learning or close variations, used without further qualification. | Machine Learning, Machine Learning Models, Natural Language Processing |
| Broad AI | Broad, non-specific mentions of artificial intelligence or close variations, used without further qualification. | Artificial Intelligence, General Artificial Intelligence, AI Algorithm |
| Unknown or Domain-Specific Algorithms | Algorithms whose structure cannot be inferred, that are proprietary, highly specialized, or not identifiable as belonging to standard categories. | iCAMP, DeepCLIP, Loihi, Custom algorithms |

The following prompt was Applied once per keyword–category pair. {keyword} is the focal term; {sentence} is the surrounding context; {definition}, {signals}, and {examples} are drawn from Table S2 for the category being evaluated.

*SI Table 6. LLM Prompt — Stage 3: Algorithm Category Classification*

Overview: You are an expert annotator, classifying the role played by a tool in a specific part of a research proposal as described by a sentence.

tool: {keyword}
sentence: {sentence}

Task: Determine which research usage(s) the tool is used for in the SENTENCE above.
Research usages: {stage_defs}

Rules (follow strictly):
- Consider ALL usages jointly.
- Assign ONLY usages that are explicitly mentioned in the sentence.
- Prefer the most specific usage(s).
- Do NOT infer unstated or downstream uses.
- If no usage clearly applies, assign all scores = 0.
- If the score for a usage is 0, leave its rationale blank.

Scoring guidance:
- 1.0  = the usage is EXPLICITLY AND DIRECTLY mentioned.
- 0.75 = the usage is EXPLICITLY AND DIRECTLY stated, but only partially aligned with the definition.
- 0.5  = the usage is EXPLICITLY OR DIRECTLY stated and only partially aligned with the definition.
- 0.25 = the usage is IMPLIED or WEAKLY ALIGNED with the definition.
- 0.0  = the usage is not mentioned, or is misaligned.

Use the FULL score range when appropriate. Reserve 1.0 ONLY when {keyword} is EXPLICITLY AND DIRECTLY mentioned in the defined usage. If you are IMPLYING the usage, give 0.25.

Return STRICT JSON only:
{"scores": {"stage_name": 0.0–1.0},
 "rationales": {"stage_name": "Brief explanation"}}
Return ONLY valid JSON. Do not include explanations.

SI Fig. 3 shows the most observed keywords in each algorithm category.

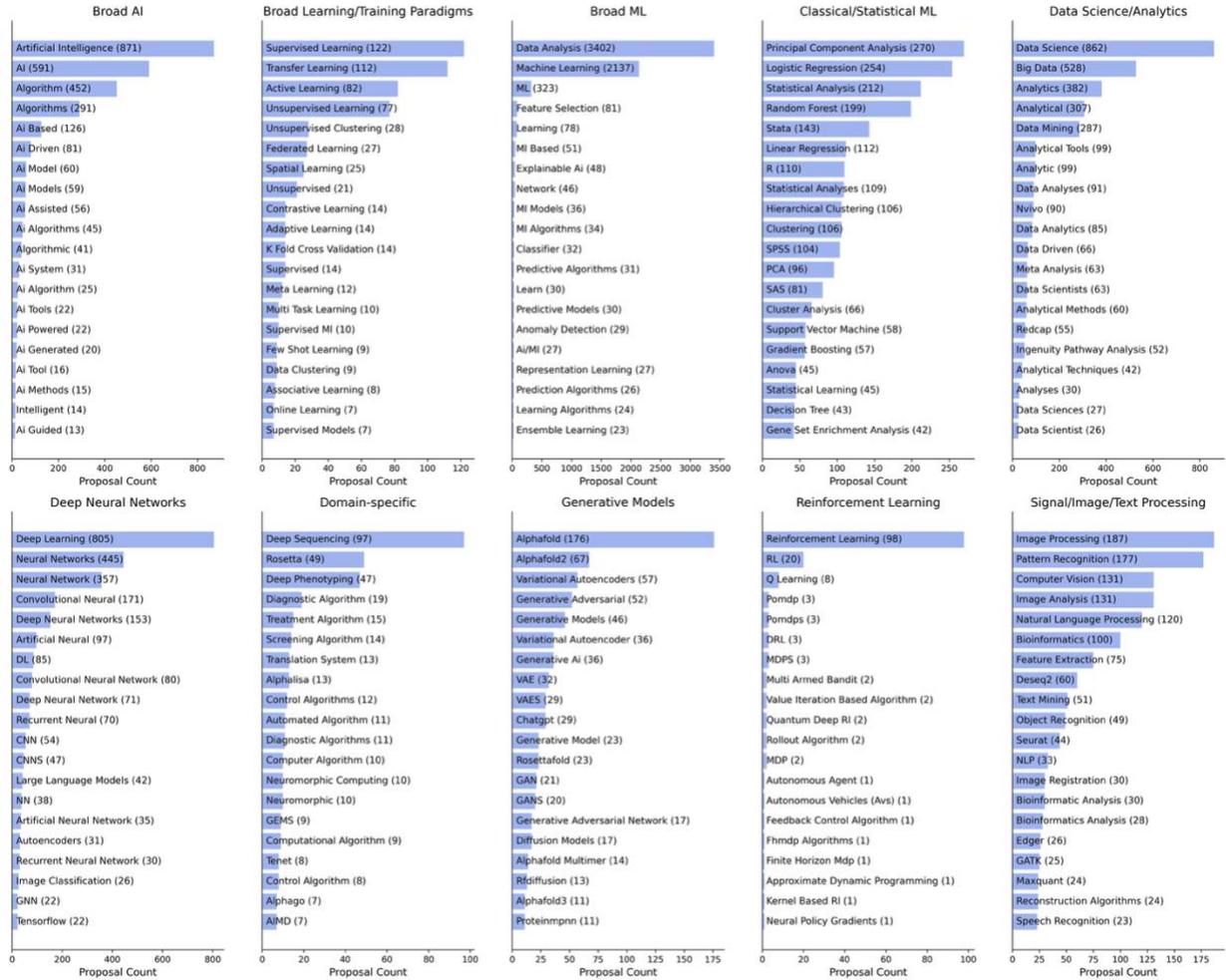

*SI Figure 3. The 10 Most Frequently Observed Keywords in Each Algorithm Category.*

## Adoption Patterns

Aggregating adoption patterns based on the share of keywords across proposals in each year, we replicate the pattern shown in Fig. 1, which were aggregated at the proposal-level.

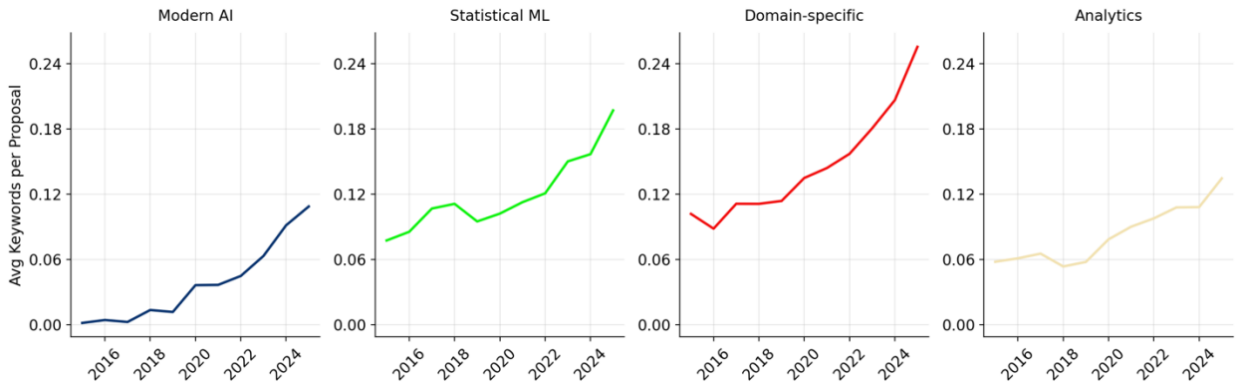

*SI Figure 4. Figure. Average Number of Unique AI-Related Keywords per Proposal by Algorithm Category, Over Time. Each panel corresponds to an algorithm category identified in proposal texts. Lines show the average number of unique keywords per submitted research proposal in a given application year. Keyword counts are normalized by the total number of research proposals submitted that year. Patterns complement the main Fig. 1 (b) that shows the fraction of proposals using an algorithm category.*

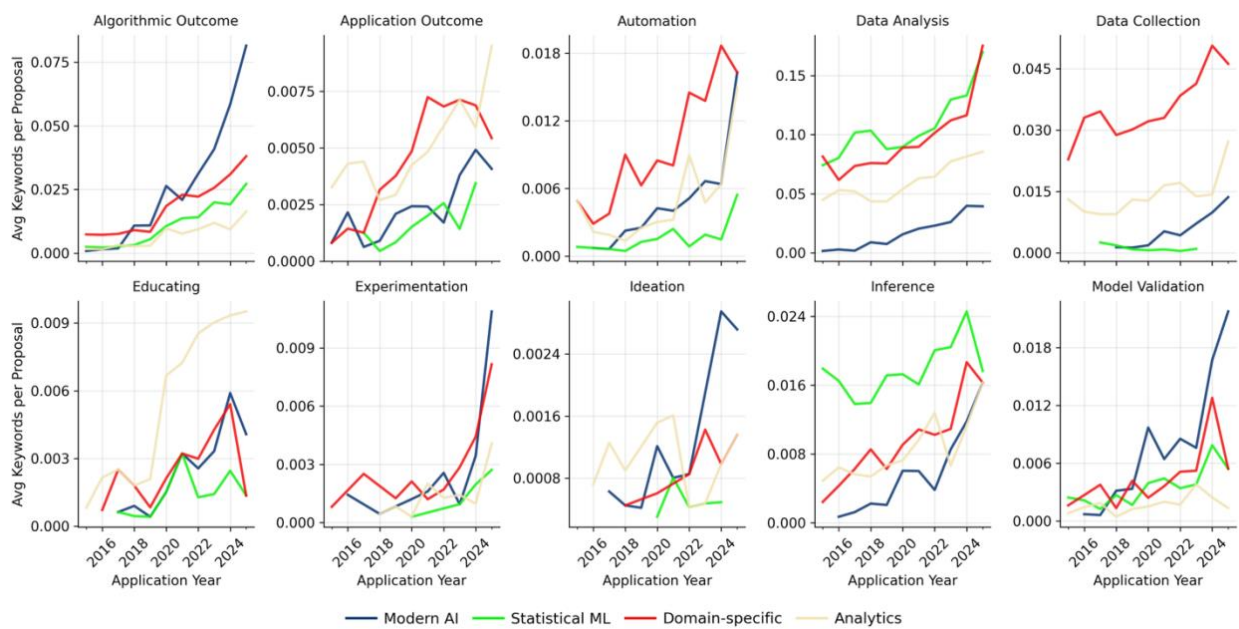

***SI Figure 5. Figure. Average Number of Unique AI-Related Keywords per Proposal by Usage and Algorithm Category, Over Time.*** *Each panel corresponds to a usage category identified in proposal texts. Lines show the average number of unique keywords belonging to each algorithm category (Modern AI, Statistical ML, Domain-specific, Analytics) per submitted research proposal in a given application year. Keyword counts are normalized by the total number of research proposals submitted that year. Patterns complement the main Fig. 1 (c) that shows the fraction of proposals using an algorithm category for a specific use case. Specifically, this aggregation approach reveals modern AI in prominent in experimentation, ideation, and development (algorithmic outcome and model validation, which is a part of validating the algorithm).*

## Heterogeneity across AI Adopters

Finally, we examine heterogeneity in the association between AI adoption and research characteristics across different classes of AI methods. Our empirical analysis reveals gender differences in AI adoption, particularly, the modern AI. Moreover, modern AI adopters are marginally younger, and yet appear to be more established before deciding to adopt, consistent with AI's role in establishing early-career scholars (Hao et al., 2023).

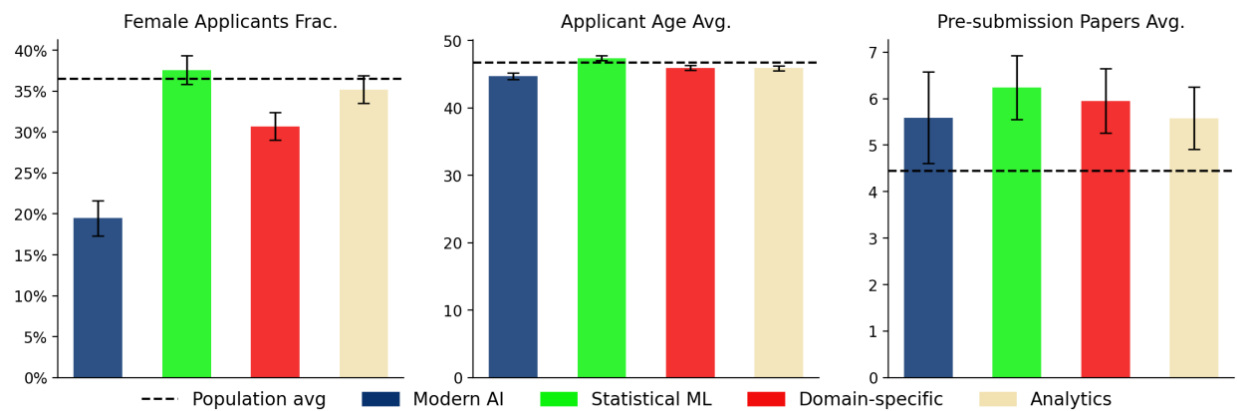

***SI Figure 6. Demographic Information and AI Adoption.*** *The figure shows the average (and confidence intervals) for the fraction of female main applicants, average age of the main applicant, and the average number of main applicant's publications prior to submitting the proposals across algorithm categories.*

# Effect of Modern AI versus Other Algorithm Groups
## Joint Estimates

Here, we use specifications like the main Fig. 2c-d but include binary variables for the use of each algorithm category, offering a joint estimate which helps isolate the distinct effect of these methods on proposal characteristics. The only significance difference from the main estimates is that modern AI is associated with lower total requested budget amounts.

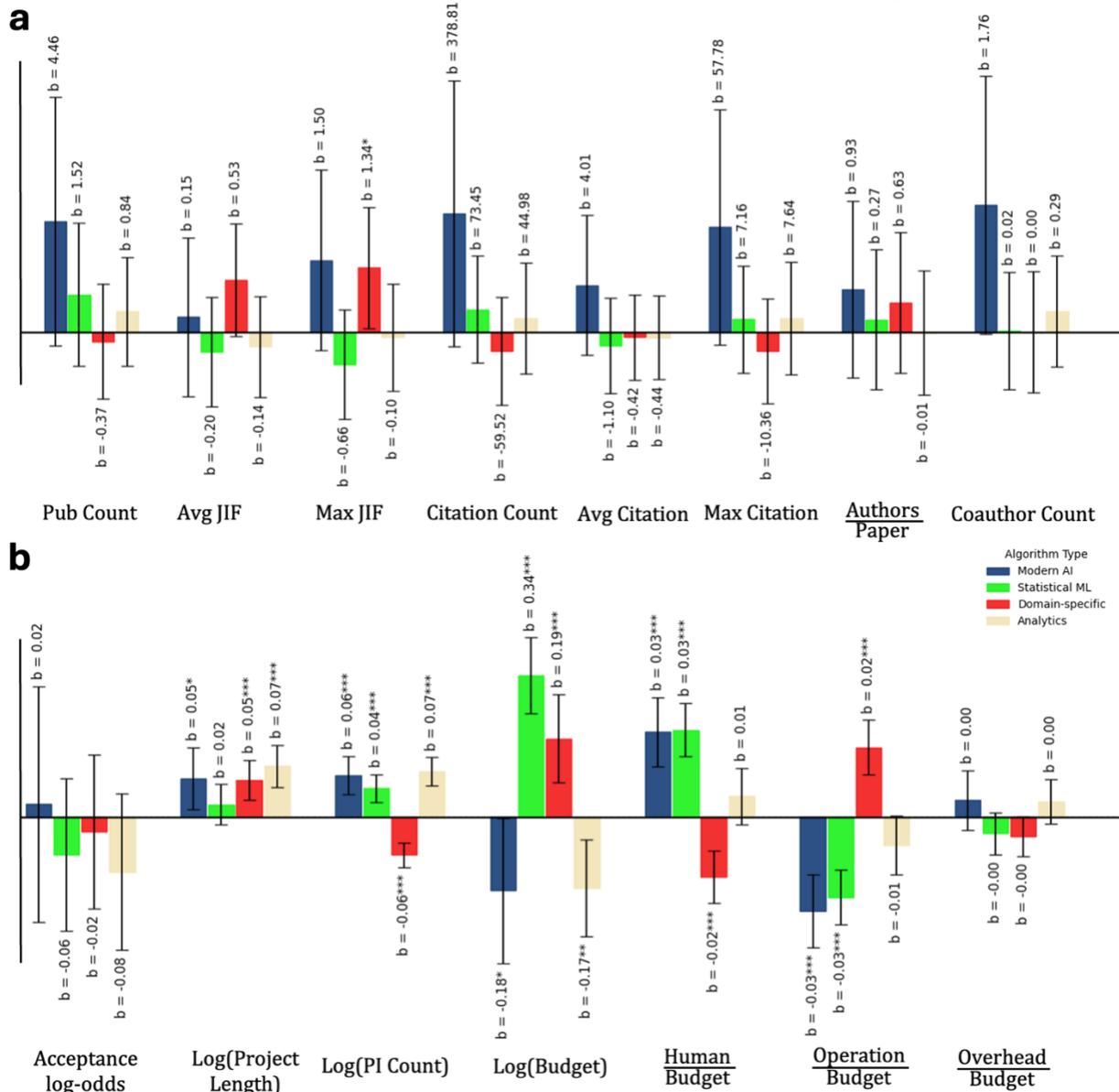

**SI Figure 7. Joint Estimation of the Association between Algorithm Groups and Variables of Interest.** *Regressions follow specifications of the main Fig. 2c-d but include binary variables for the use of each algorithm category. Isolating these associations, the minute positive coefficients on scientific outcomes are no longer statistically significant. However, the joint estimates reveal that modern AI is associated with a lower total requested budget, documenting the only major difference to the estimates of main Fig. 2d. Bar lengths are scaled as a function of coefficients scaled by the variation in the dependent variable.*

# Regressions based on Matched Sample of Modern AI and Non-AI Proposals

To further address confounding beyond regression-based controls, we also use proposal text to construct a semantically matched sample of AI-enabled and conventional proposals. Any proposal that uses any type of algorithm is excluded from the pool of possible matched proposals. This matching procedure allows us to compare projects that address closely related scientific problems while differing in their use of AI. SI Table 7 shows the estimates equivalent to those in the main Fig 2d (input and process estimates) and SI Table 8 shows estimates equivalent to the main Fig 2c (scientific outputs for the funded proposals).

*SI Table 7. Regression Estimates of Input and Process Variables on AI-enabled Proposals based on Semantically Matched Sample*

| VARIABLES | (1) Acceptance Log Odds | (2) log(Project Length) | (3) Log(Number of PIs) | (4) log(Requested Budget) | (5) Human Capital /Budget | (6) Human Costs /Budget | (7) Overhead /Budget |
|---|---|---|---|---|---|---|---|
| Treated (Modern AI) | -0.350*** | 0.111*** | 0.143*** | -0.134 | 0.022*** | -0.020*** | -0.001 |
|  | (-0.537 - -0.163) | (0.067 - 0.154) | (0.115 - 0.172) | (-0.296 - 0.027) | (0.012 - 0.032) | (-0.031 - -0.009) | (-0.006 - 0.003) |
| Age Groups: 35-49 | -0.447*** | -0.028 | 0.060*** | 0.082 | -0.011** | 0.062*** | -0.051*** |
|  | (-0.630 - -0.265) | (-0.068 - 0.011) | (0.037 - 0.083) | (-0.074 - 0.238) | (-0.021 - -0.001) | (0.050 - 0.074) | (-0.058 - -0.044) |
| Age Groups: 50-64 | -0.469*** | 0.065*** | 0.167*** | 0.297*** | -0.035*** | 0.084*** | -0.049*** |
|  | (-0.675 - -0.262) | (0.019 - 0.110) | (0.137 - 0.197) | (0.125 - 0.468) | (-0.047 - -0.022) | (0.070 - 0.098) | (-0.057 - -0.042) |
| Age Groups: >65 | -1.390*** | 0.153*** | 0.118*** | 0.406*** | 0.032*** | 0.020* | -0.052*** |
|  | (-1.808 - -0.971) | (0.050 - 0.256) | (0.072 - 0.163) | (0.203 - 0.610) | (0.011 - 0.053) | (-0.001 - 0.041) | (-0.063 - -0.041) |
| Female | -0.282*** | -0.018 | 0.032*** | -0.006 | 0.025*** | -0.013*** | -0.012*** |
|  | (-0.431 - -0.133) | (-0.053 - 0.017) | (0.010 - 0.054) | (-0.124 - 0.111) | (0.016 - 0.034) | (-0.022 - -0.004) | (-0.017 - -0.008) |
| log(Project Length) | 0.710*** |  | -0.014* | 0.293*** | 0.015*** | -0.013*** | -0.002** |
|  | (0.585 - 0.836) |  | (-0.029 - -0.001) | (0.210 - 0.375) | (0.008 - 0.021) | (-0.019 - -0.006) | (-0.004 - -0.000) |
| Number of PIs | -0.012 | -0.014 |  | -0.212*** | 0.008*** | -0.013*** | 0.005*** |
|  | (-0.074 - 0.050) | (-0.035 - 0.006) |  | (-0.318 - -0.107) | (0.003 - 0.014) | (-0.019 - -0.007) | (0.002 - 0.007) |
| log(Requested Budget) | 0.001 | 0.024*** | -0.014*** |  | -0.007*** | 0.008*** | -0.001** |
|  | (-0.019 - 0.020) | (0.015 - 0.032) | (-0.020 - -0.008) |  | (-0.009 - -0.005) | (0.005 - 0.011) | (-0.002 - -0.000) |
| Similarity to Proposals | -2.632 | 2.358*** | -0.844*** | 0.190 | -1.610*** | 1.961*** | -0.352*** |
|  | (-6.463 - 1.199) | (1.238 - 3.477) | (-1.450 - -0.238) | (-3.140 - 3.521) | (-1.852 - -1.368) | (1.688 - 2.235) | (-0.468 - -0.235) |
| Pre-proposal PI Pub Count | 0.009*** | -0.001*** | 0.003*** | 0.003 | 0.000 | 0.000 | -0.000*** |
|  | (0.005 - 0.012) | (-0.003 - -0.000) | (0.002 - 0.005) | (-0.003 - 0.008) | (-0.000 - 0.000) | (-0.000 - 0.001) | (-0.000 - -0.000) |
| Accepted Grant |  | 0.296*** | -0.004 | -0.024 | -0.020*** | 0.021*** | -0.001 |
|  |  | (0.253 - 0.339) | (-0.027 - 0.019) | (-0.171 - 0.123) | (-0.032 - -0.008) | (0.009 - 0.034) | (-0.006 - 0.004) |
| Constant | -0.514 | 0.380 | 1.268*** | 12.606*** | 2.072*** | -1.531*** | 0.459*** |
|  | (-3.836 - 2.807) | (-0.587 - 1.348) | (0.742 - 1.794) | (9.804 - 15.408) | (1.853 - 2.291) | (-1.780 - -1.283) | (0.359 - 0.559) |
|  |  |  |  |  |  |  |  |
| Observations | 7,437 | 7,813 | 7,813 | 7,813 | 6,596 | 6,596 | 6,596 |
| R-squared |  | 0.328 | 0.107 | 0.093 | 0.190 | 0.131 | 0.255 |
| Matched sample | Modern AI | Modern AI | Modern AI | Modern AI | Modern AI | Modern AI | Modern AI |
| Year FE | yes | yes | yes | yes | yes | yes | yes |
| Cluster | matched samples | matched samples | matched samples | matched samples | matched samples | matched samples | matched samples |

Robust ci in parentheses
*** p<0.01, ** p<0.05, * p<0.1

*SI Table 8. Regression Estimates of Outcome Variables on AI-enabled Proposals based on Semantically Matched Sample*

| VARIABLES | (1) Pub Count | (2) Avg. JIF | (3) Max JIF | (4) Citations Count | (5) Avg. Citations | (6) Max citations | (7) Authors/ Paper | (8) Coauthor Count |
|---|---|---|---|---|---|---|---|---|
| Treated (Modern AI) | 8.499* | -0.110 | 3.273** | 544.915 | 6.005 | 147.424 | 0.310 | 2.696* |
|  | (-0.628 - 17.626) | (-1.443 - 1.224) | (0.169 - 6.378) | (-335.382 - 1,425.212) | (-19.284 - 31.294) | (-58.537 - 353.385) | (-4.348 - 4.967) | (-0.468 - 5.860) |
| Age Groups: 35-49 | -0.255 | -0.028 | -1.373 | -175.148 | -0.751 | 13.190 | 3.749*** | 0.512 |
|  | (-5.906 - 5.397) | (-1.063 - 1.007) | (-3.750 - 1.004) | (-579.630 - 229.334) | (-16.704 - 15.201) | (-101.413 - 127.792) | (0.976 - 6.522) | (-1.546 - 2.571) |
| Age Groups: 50-64 | -0.107 | -1.306* | -0.914 | 6.982 | -10.158 | 153.833 | -2.487 | 2.355* |
|  | (-6.814 - 6.599) | (-2.674 - 0.062) | (-4.083 - 2.256) | (-889.028 - 902.991) | (-31.786 - 11.470) | (-151.136 - 458.801) | (-6.243 - 1.270) | (-0.052 - 4.762) |
| Age Groups: = >65 | -3.476 | -2.647** | -7.577** | -1,319.955* | 117.585 | -174.848 | -8.608** | -1.567 |
|  | (-29.211 - 22.259) | (-5.153 - -0.140) | (-14.229 - -0.925) | (-2,721.711 - 81.801) | (-183.653 - 418.823) | (-555.311 - 205.615) | (-15.489 - -1.727) | (-9.498 - 6.364) |
| Female | -4.337* | -1.357** | -2.255** | 356.934 | 8.319 | 81.941 | 2.058 | -1.322* |
|  | (-9.038 - 0.364) | (-2.422 - -0.292) | (-4.426 - -0.084) | (-317.514 - 1,031.383) | (-12.131 - 28.768) | (-93.171 - 257.053) | (-1.939 - 6.055) | (-2.817 - 0.173) |
| Number of PIs | -0.423 | 0.889*** | 0.736 | 802.868** | 26.602*** | 144.861** | 7.841*** | -0.043 |
|  | (-3.618 - 2.772) | (0.334 - 1.443) | (-0.422 - 1.894) | (82.023 - 1,523.713) | (7.613 - 45.591) | (15.079 - 274.643) | (3.745 - 11.938) | (-1.090 - 1.004) |
| log(Requested Budget) | 6.567*** | 0.684*** | 2.556*** | 250.444** | -0.050 | 39.981* | 0.000 | 2.835*** |
|  | (3.129 - 10.005) | (0.361 - 1.008) | (1.360 - 3.751) | (54.498 - 446.391) | (-6.415 - 6.314) | (-2.354 - 82.316) | (-1.220 - 1.220) | (1.276 - 4.394) |
| Similarity to Proposals | 396.173*** | 53.005*** | 289.803*** | 34,213.225*** | 1,040.433*** | 8,503.321*** | 363.816*** | 114.004*** |
|  | (247.789 - 544.557) | (26.669 - 79.342) | (217.470 - 362.136) | (18,818.304 - 49,608.145) | (513.138 - 1,567.728) | (4,119.008 - 12,887.634) | (234.445 - 493.187) | (66.299 - 161.709) |
| Pre-proposal PI Pub Count | 0.702** | 0.058*** | 0.184*** | 44.614*** | 0.867*** | 3.943** | 0.221*** | 0.249** |
|  | (0.136 - 1.267) | (0.029 - 0.087) | (0.115 - 0.253) | (10.823 - 78.405) | (0.324 - 1.410) | (0.450 - 7.436) | (0.083 - 0.358) | (0.031 - 0.466) |
| Constant | -402.493*** | -48.125*** | -269.044*** | -28,075.686*** | -767.841*** | -7,117.690*** | -294.133*** | -129.799*** |
|  | (-537.839 - -267.146) | (-72.653 - -23.597) | (-333.374 - -204.715) | (-40,456.733 - -15,694.639) | (-1,219.919 - -315.764) | (-10,621.236 - -3,614.143) | (-406.284 - -181.983) | (-176.194 - -83.405) |
| Observations | 642 | 592 | 592 | 642 | 396 | 396 | 642 | 641 |
| R-squared | 0.259 | 0.132 | 0.278 | 0.253 | 0.290 | 0.234 | 0.355 | 0.271 |
| Year FE | yes | yes | yes | yes | yes | yes | yes | yes |
| Cluster | Matched Groups | Matched Groups | Matched Groups | Matched Groups | Matched Groups | Matched Groups | Matched Groups | Matched Groups |

Robust ci in parentheses
*** p<0.01, ** p<0.05, * p<0.1

## Regressions based on Ordered Categories of Modern AI Use

Here, we use specifications like the main Fig. 2c-d but include a d variable for the use of modern AI in proposals. This categorical variable distinguishes proposals that do not use modern AI (control), proposals with modern AI-related keyword count below the annual median, and those above the median. Results are consistent with the main Figure.

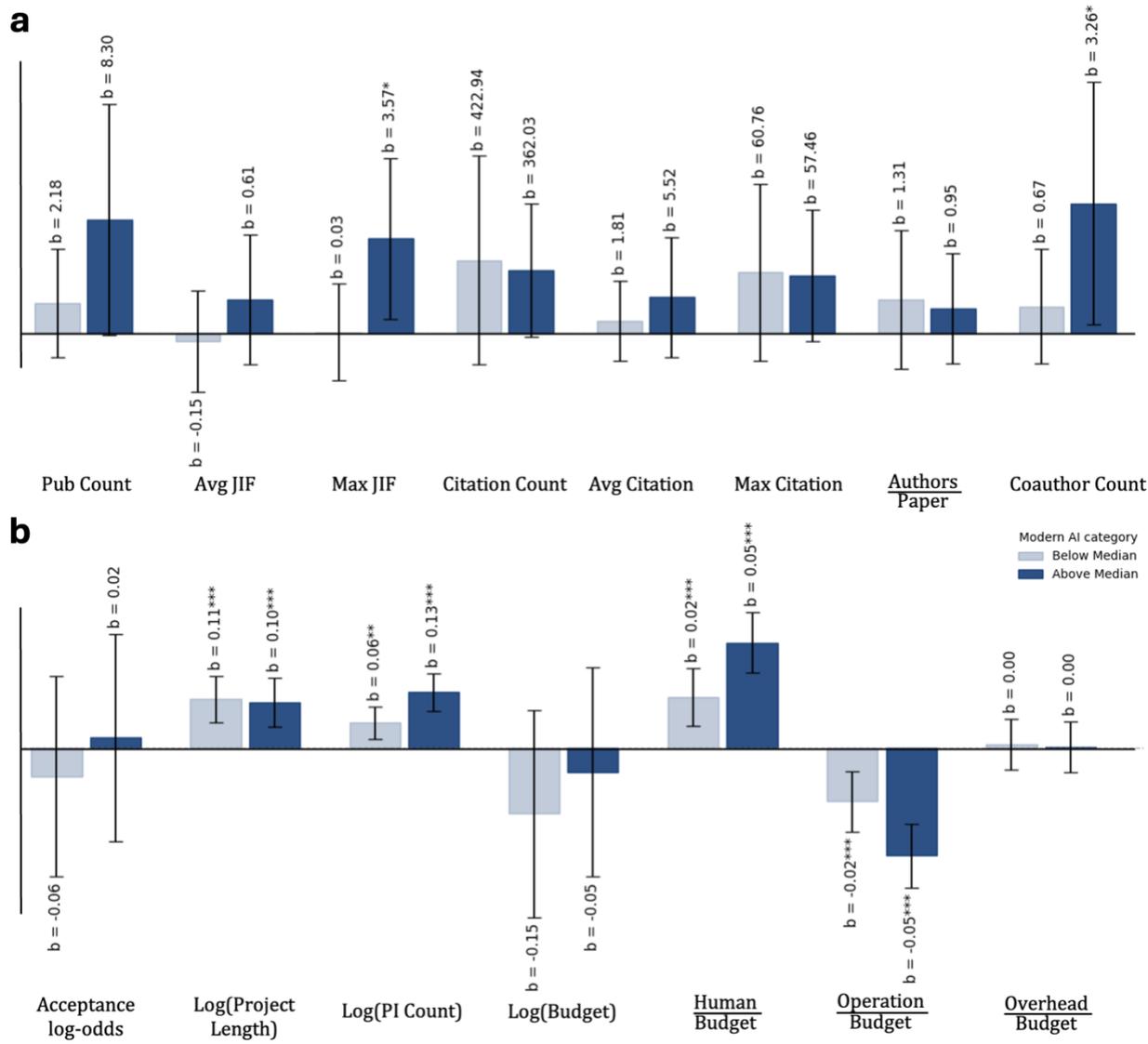

*SI Figure 8. Regression Analysis of the Association of AI based on an Ordered Categorical Measure.* The specifications replicate the regressions in the main Fig. 2c-d but include a categorical variable for the use of modern AI in proposals. Results are consistent with estimates in the main Figure.

## The Task Composition of AI-enabled and Non-AI Proposals

Here, we analyze differences in the prevalence of specific tasks, ranking them by the difference in frequency between AI-enabled and conventional proposals. This approach allows us to characterize whether AI adoption is associated with substitution across tasks or the expansion of task sets within projects.

Three confidence thresholds (80%, 85%, 90%) were evaluated. Raising the threshold from 80% to 90% reduced the number of proposals with at least one extracted task from 20,093 to 17,243 but left the rank ordering of proposals by task count and the statistical contrasts between AI and non-AI projects substantively unchanged (Mann-Whitney and Kolmogorov-Smirnov tests stable across thresholds). The 80% threshold was adopted for the primary analysis to maximize coverage. Manual inspection of extracted tasks confirmed face validity:

examined proposals yielded task statements commensurate with the described research activities, and matched treated-control pairs, including a targeted check of AlphaFold-related proposals and their matched controls, produced task lists that differed in interpretable ways.

SI Fig. 9 and SI Fig. 10 show the top 15 tasks primarily associated with AI-enabled and Non-AI proposals, respectively. Whereas Most tasks done primarily by AI-enabled proposals are related to analyzing data, or developing models, the tasks in which Non-AI proposals have higher fractions are contextually relevant to both sets of proposals. Therefore, the difference between the fractions of Non-AI and AI-enabled proposals undertaking these tasks are relatively smaller, underlying the expansion, rather than the substitution effect observed in the main Fig. 4.

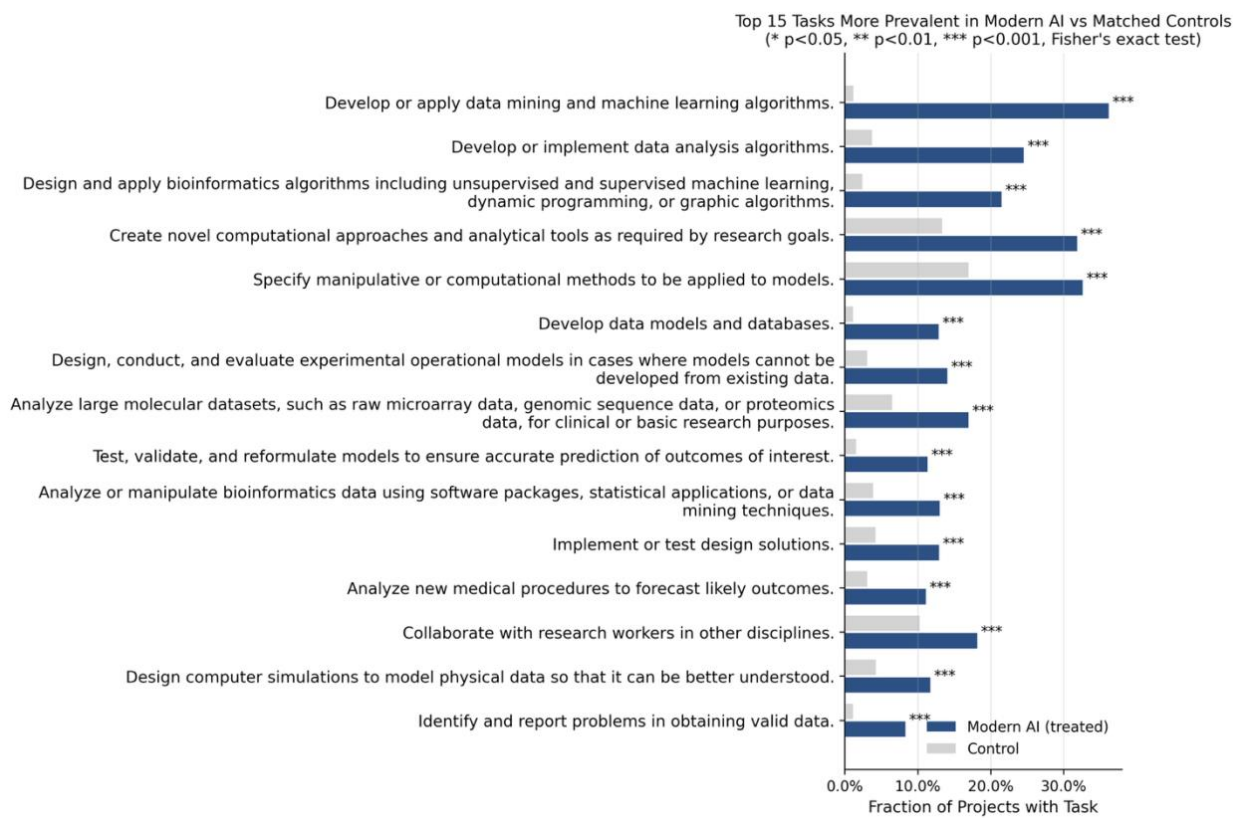

*SI Figure 9. Top 15 Tasks Primarily Undertaken by AI-enabled Proposals. These tasks are selected by comparing the first quantifying the fraction of AI-enabled and Non-AI proposals undertaking the task, and then ordering them based on the difference between the two fractions. Most tasks done primarily by AI-enabled proposals are related to analyzing data, or developing models.*

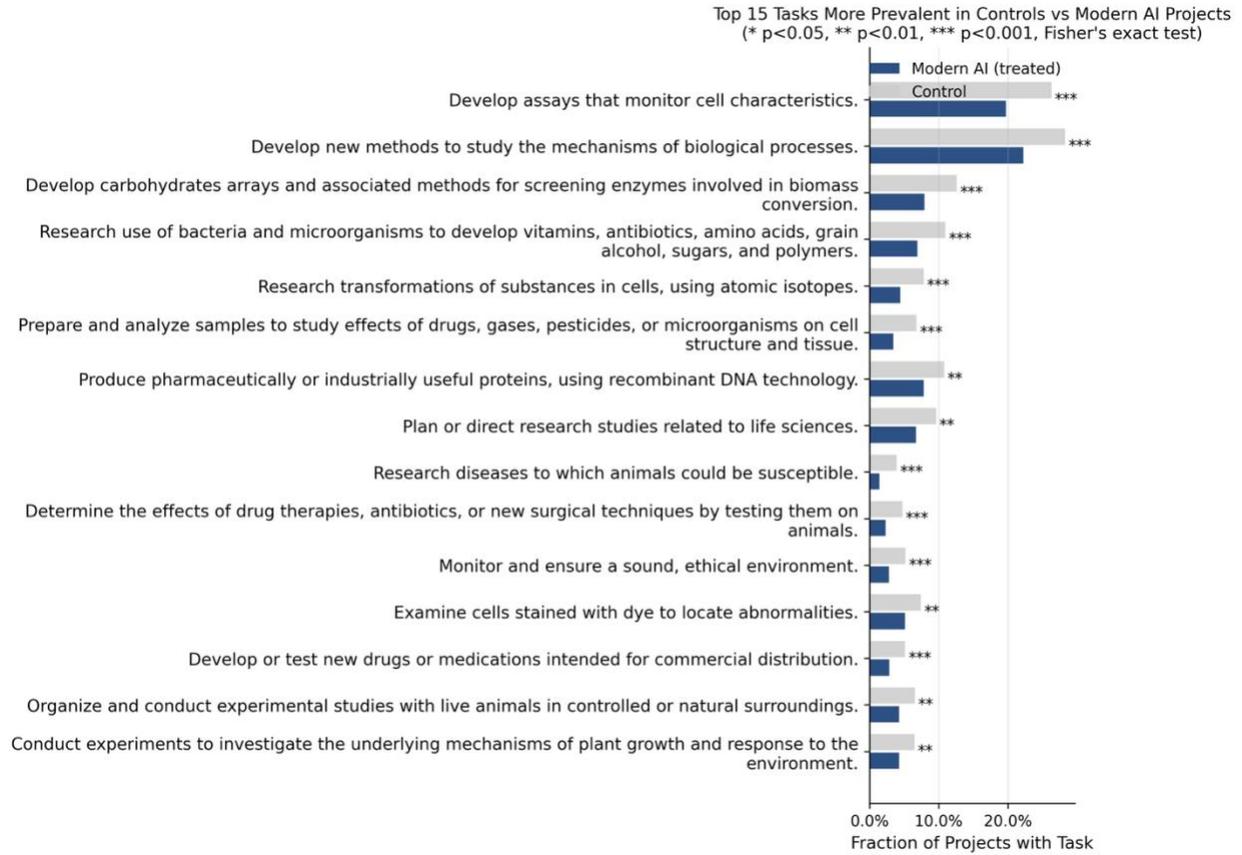

**SI Figure 10. Top 15 Tasks Primarily Undertaken by Non-AI Proposals.** *These tasks are selected by comparing the first quantifying the fraction of Non-AI and AI-enabled proposals undertaking the task, and then ordering them based on the difference between the two fractions. Most of the resulting tasks are contextually relevant to both sets of proposals, Therefore, the difference between the fractions of Non-AI and AI-enabled proposals undertaking these tasks are relatively smaller, even when the difference is statistically significant. The pattern is consistent with the expansion, rather than the substitution effect observed in the main Fig. 4.*

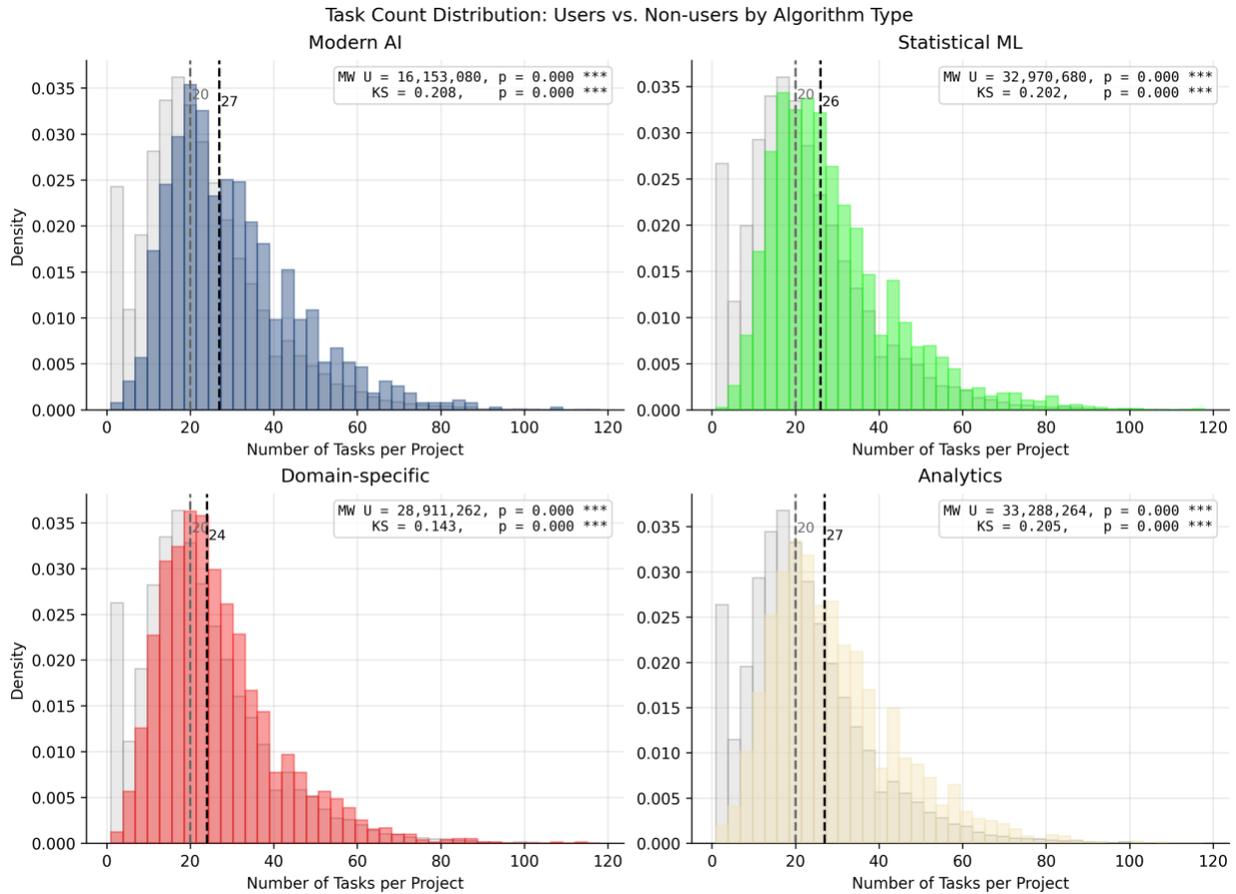

**SI Figure 11. Task Counts for Different Algorithm Categories Compared to the Baseline of Non-AI Proposals.**

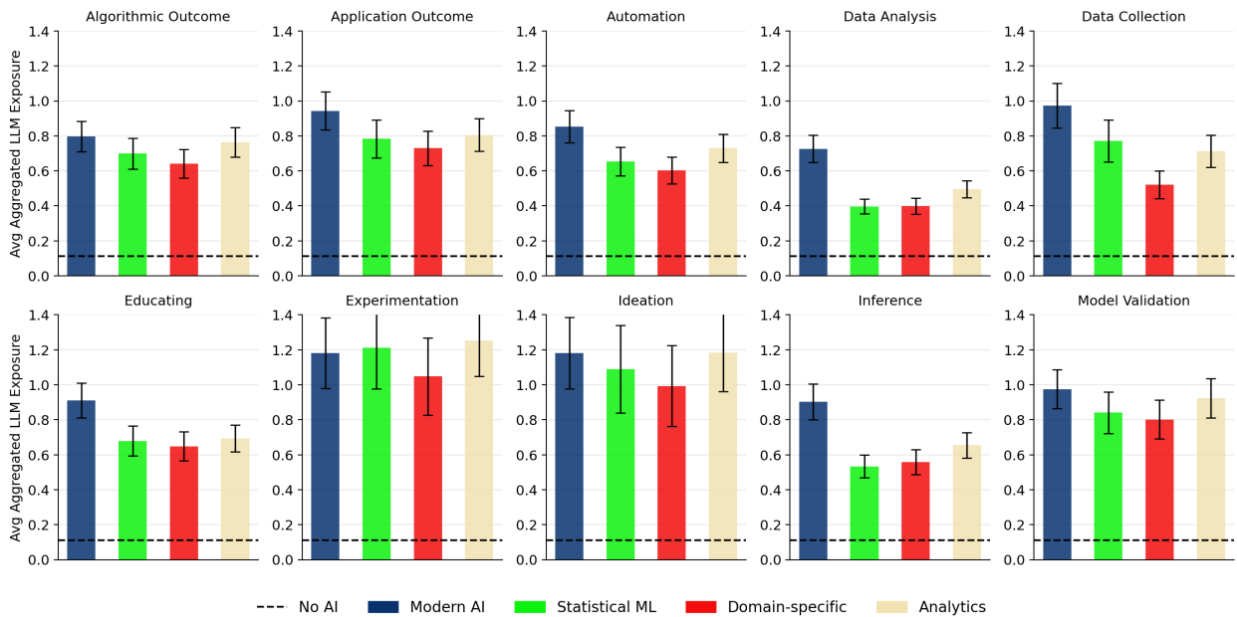

**SI Figure 12. Aggregated LLM Exposure across Usage and Algorithm Categories.** *To estimate the LLM exposure for each proposal submitted after 2022, we first match extracted tasks to the newly released Anthropic LLM exposure scores. These exposure scores are then summed at the level of a proposal, aggregated LLM exposure, to measure the amenability of the proposal activities to be performed by LLMs. The figure expands the comparison in main Figure 5 to all algorithm categories.*